\documentclass[aps,prb,reprint,groupedaddress
]{revtex4-1}
\usepackage{bm,amssymb,graphicx}
\def\be{\begin{equation}}  
\def\ee{\end{equation}}  
\def\ba{\begin{eqnarray}}  
\def\ea{\end{eqnarray}}  
\def\bc{\begin{center}}  
\def\ec{\end{center}}  
\def\p{\partial}  
\begin{document}


\title{Microwave-induced ``zero-resistance'' states and second-harmonic generation in an ultraclean two-dimensional electron gas }


\author{S.~A.~Mikhailov}
\affiliation{Institute of Physics, University of Augsburg, D-86161 Augsburg, Germany}



\date{\today}

\begin{abstract}
Microwave-induced resistance oscillations (MIRO) and ``zero-resistance'' states (ZRS) were discovered in ultraclean two-dimensional electron systems in 2001--–2003 and have attracted great interest from researchers. A comprehensive theory of these phenomena was developed in 2011: It was shown that all experimentally observed dependencies can be naturally explained by the influence of the ponderomotive forces which arise in the near-contact regions of the two-dimensional electron gas under the action of microwaves. Now we show that the same near-contact physical processes should lead to another nonlinear electrodynamic phenomenon -- the second-harmonic generation. We calculate the frequency, magnetic field, mobility, and power dependencies of the second-harmonic  intensity and show that it can be as large as $\gtrsim 0.5$ mW/cm$^2$ under realistic experimental conditions. A part of this paper is devoted to a further discussion of the MIRO/ZRS phenomena: we explain how the ponderomotive-force theory explains different experimental details, including those which were not known in 2011, and critically discuss alternative theories.
\end{abstract}


\maketitle


\section{introduction}

\begin{figure}
\includegraphics[width=8.5cm]{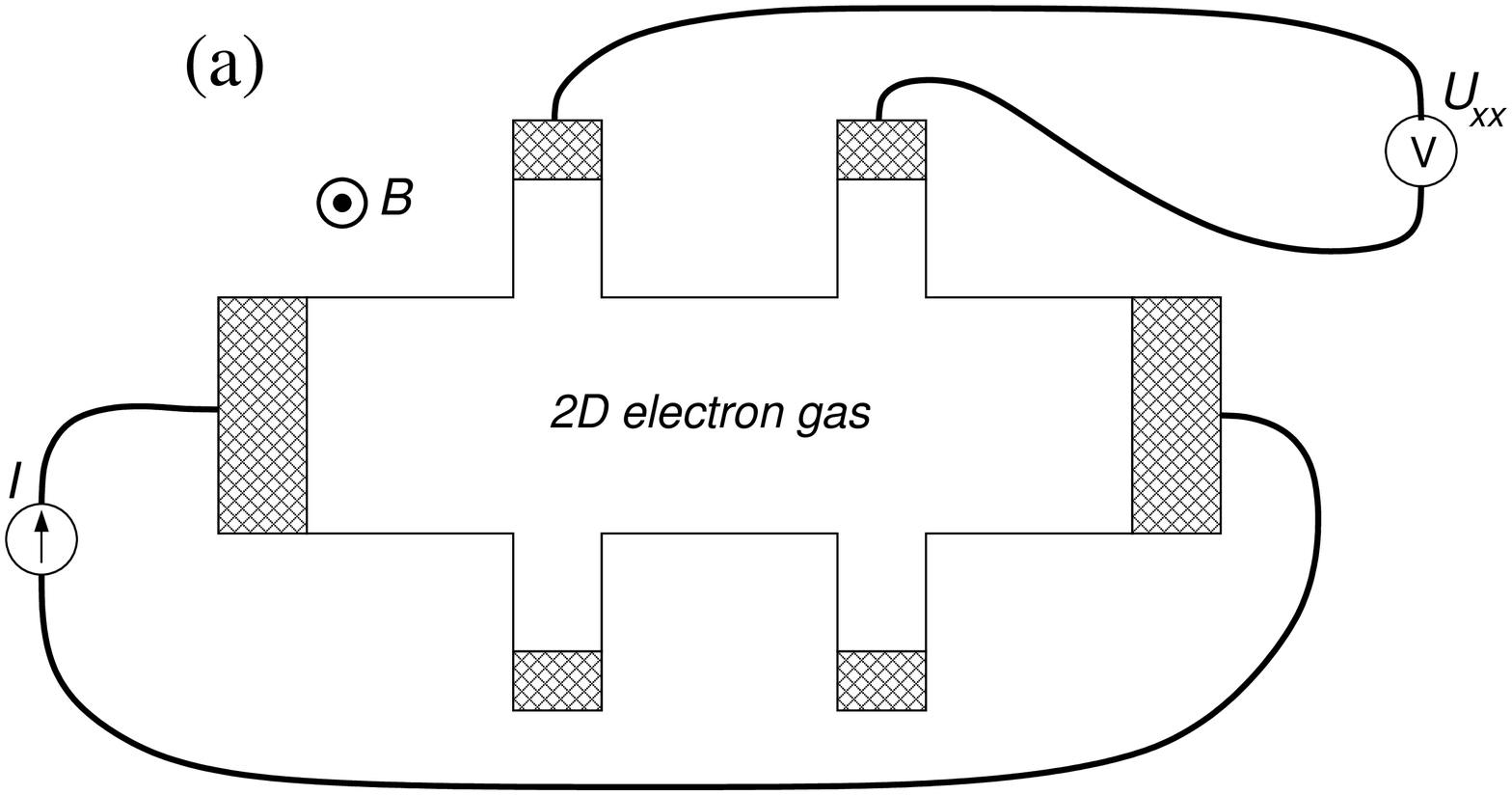}\\ \vspace{5mm}
\includegraphics[width=8.5cm]{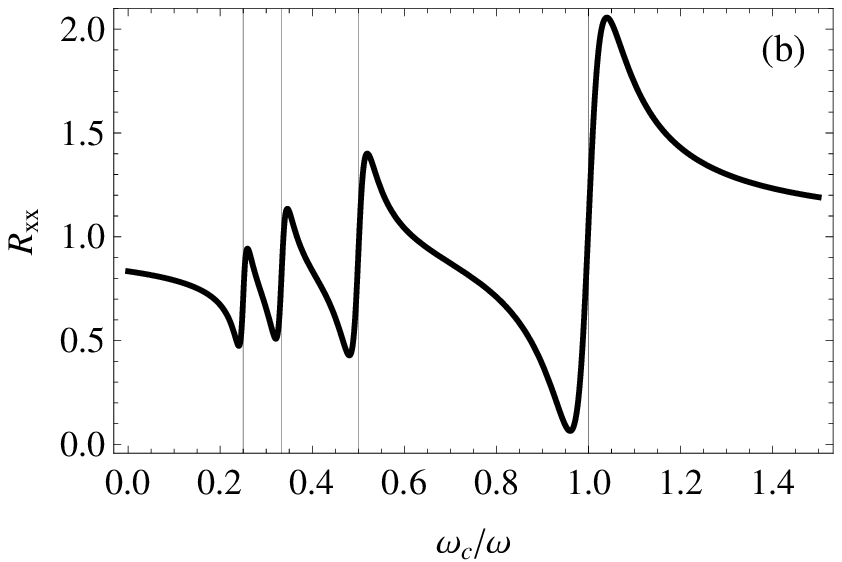}
\caption{\label{fig:geometry} (a) The geometry of a standard Hall bar in which the MIRO/ZRS effects have been observed\cite{Zudov01,Ye01,Mani02,Zudov03}. A dc current flows from the left to right contact (shaded areas), the voltage is measured between  the side contacts, and the resulting (longitudinal) resistance is $R_{xx}=U_{xx}/I$. The sample is placed in a perpendicular magnetic field $B$ and is irradiated by microwaves with the frequency $\omega=2\pi f$.  (b) A qualitative dependence of the experimentally observed Hall-bar resistance as a function of the magnetic field $B$; here $\omega_c/\omega=eB/mc\omega$. At the points $\omega=k\omega_c$, i.e. $\omega_c/\omega=1/k$, $k=1,2,\dots$, shown by thin vertical lines, the measured microwave photoresistance coincides with the dark magnetoresistance. Right from these points the photoresistance is substantially larger than the dark one (by a factor of $7-10$ is real experiments, see e.g. Ref. \cite{Mani02}). Left from the points $\omega_c/\omega=1/k$ the photoresistance decreases as compared to the dark one and can be suppressed down to zero if the microwave power is sufficiently large.}
\end{figure}

The microwave-induced resistance oscillations (MIRO) and ``zero-resistance'' states (ZRS) were discovered in very-high-electron-mobility two-dimensional (2D) electron systems in GaAs-AlGaAs quantum wells in 2001---2003 \cite{Zudov01,Ye01,Mani02,Zudov03}. The magnetoresistance $R_{xx}=U_{xx}/I$ of a 2D electron gas, measured between the side contacts to a Hall-bar sample, see Figure \ref{fig:geometry}(a), demonstrated very large oscillations around the dark value $R_{xx}^0$ under the action of microwaves, Figure \ref{fig:geometry}(b). These oscillations are governed by the parameter $\omega/\omega_c$, where $\omega$ and $\omega_c=eB/mc$ are the microwave and cyclotron frequencies, respectively, $B$ is the external magnetic field, perpendicular to the 2D electron-gas plane, $e$ and $m$ are the charge and effective mass of 2D electrons, and $c$ is the velocity of light. If the microwave power is sufficiently strong ($\gtrsim 1$ mW/cm$^2$), maxima of the measured magnetoresistance $R_{xx}$ can be $7-10$ times larger than the dark $R_{xx}$ values, while the minima demonstrate apparently vanishing $R_{xx}$ (or $U_{xx}$) values and were therefore called ``zero-resistance states''. The effect is observed not only around the fundamental cyclotron frequency $\omega=\omega_c$, but also around  harmonics $\omega=k\omega_c$, with the harmonic index $k$ up to $\sim 10$; the amplitudes of the $R_{xx}$ oscillations decrease with $k$. The MIRO/ZRS effect is seen in low magnetic fields ($B\lesssim 0.5$ T) when the number of occupied Landau levels is about $50-100$. Practically no influence of the microwave radiation on the Hall resistance $R_{xy}$ was observed in the MIZRS regime (no microwave induced Hall plateaus which could be expected by analogy with the quantum Hall effect\cite{Klitzing80}).

The vanishing resistance (which reminded the superconductivity, Ref. \cite{Mani02}) and a  similarity of the MIZRS phenomenon to the quantum Hall effect attracted great interest from researchers. A large number of theoretical scenarios (e.g. Refs. \cite{Durst03,Lei03a,Ryzhii03a,Koulakov03,Dmitriev03,Dmitriev04,Dmitriev05,Vavilov04}) claiming to explain the observed $R_{xx}$-oscillations were published in 2003--2005, but no one of them was completely convincing. A great puzzle of the MIRO/ZRS effects consisted in the fact that they were observed under the conditions 
\be 
\hbar/\tau\ll T\simeq \hbar\omega_c\lesssim \hbar\omega\ll E_F,
\label{class_cond}
\ee
where the classical physics had to be valid (here $E_F$ is the Fermi energy, $T$ is the temperature, and $\tau\equiv\tau_{tr}$ is the momentum relaxation time), but the only classical effect which seemed to be relevant 
was the resonant growth of the magnetoresistance at the cyclotron (in a finite-size sample -- at the magnetoplasmon) frequency due to the Joule heating of electrons. The magnetoplasmon resonance was indeed observed in the microwave magnetoresistance experiments \cite{Vasiliadou93} a decade before the discovery of the MIZRS. The Joule heating resonance \cite{Vasiliadou93}, however, had a standard Lorenzian shape, was weak (a few percent of the dark value) and was observed in samples with  one order of magnitude lower mobility $\mu\simeq 10^6$ cm$^2$/Vs. In the ultraclean samples of Refs. \cite{Mani02,Zudov03} ($\mu\gtrsim 10^7$ cm$^2$/Vs) the huge oscillations of $R_{xx}$ around the dark value, unexplainable within a simple classical approach, were seen instead. 

The absence of a simple and reasonable classical explanation of the MIRO/ZRS phenomena led us to the idea \cite{Mikhailov03c} that they can be due to some physical processes not in the bulk, but near the edge of the sample (although the specific mechanism proposed in that preprint was incorrect, it first pointed out the possibility of a non-bulk mechanism of the MIRO effect; all published in 2003--2004 theories assumed the bulk origin of MIRO). A crucial experiment which answered the question, whether the MIRO have the bulk or edge origin, was performed by Smet in 2005, Ref. \cite{Smet05}. In this experiment  the MIRO/ZRS effect was studied under the influence of the right and left {\em circularly} polarized sub-terahertz radiation. It was shown that, in contrast to the absorption (a bulk effect), which demonstrated a strong dependence on the circular polarization sense,  MIROs turned out to be completely insensitive to it. This result ruled out all bulk scenarios of the MIRO/ZRS effect \cite{Durst03,Lei03a,Ryzhii03a,Koulakov03,Dmitriev03,Dmitriev04,Dmitriev05,Vavilov04}. 

After that the theoretical activity in the MIRO/ZRS field decreased significantly. Only the group of Dmitriev et al. continued to publish papers on MIRO/ZRS (more than ten after 2005, for a recent review see \cite{Dmitriev12})  claiming that their theory, based on the so called inelastic-scattering mechanism, explains the effect, in spite of evident contradictions of their results with experimental data and a critique in the literature \cite{Mikhailov04a}. Another huge activity was started by I\~narrea, who produced more than twenty papers after his first publication \cite{Inarrea05} in 2005. In 2009 Chepelianskii et al. proposed a mechanism related to the edge of the 2D gas  \cite{Chepelianskii09}. In spite of this the true origin of the MIRO/ZRS effects remained unclear. Moreover, new experiments added new puzzles: for example, Yang et al. \cite{Yang06} showed that the MIRO effect is strongly influenced (suppressed) by a moderate ($\lesssim 1$ T) {\em parallel} magnetic field, Willett et al. observed {\em negative} values of $R_{xx}$ in some experiments, and very recently Dai with coauthors \cite{Dai10,Dai11} and then Hatke et al. \cite{Hatke11a,Hatke11b} observed a huge spike at the magnetic field corresponding to the double cyclotron harmonic $\omega=2\omega_c$. 

In 2011 an explanation for this mysterious phenomenon was found \cite{Mikhailov11a}. It turned out that {\em all} experimentally observed facts (known to us that time) can be explained if to consider \textit{the influence of contacts} and contact wires on the experimentally measured potential difference between the side contacts $U_{xx}$, Figure \ref{fig:geometry}a. Briefly (for details see Ref. \cite{Mikhailov11a} and Section \ref{subsec:mirozrs} below), the metallic contact wires, which are inevitable in the resistance measurements, serve as antennas focusing the microwave radiation in the near-contact regions of the 2D gas, like the atmospheric electric field is focused near the tops of the lightning rods during a thunderstorm. This concentrated near the contacts, strong and strongly inhomogeneous ac electric field acts on 2D electrons by a ponderomotive force $\bm F_{pm}\propto \sigma''_{xx}(\omega,q) \bm\nabla\left(E^2(\bm r,t)\right)$, which repels electrons from or attracts them to the contacts dependent on the sign of the imaginary part $\sigma''_{xx}$ of the dynamic nonlocal conductivity $\sigma_{xx}(\omega,q)$; here $q$ is a characteristic wave vector of the inhomogeneous near-contact electric field, see \cite{Mikhailov11a}. In particular, in the ZRS regime electrons are so strongly repelled from the contacts that depletion regions are formed near them, the contacts turned out to be electrically isolated from the bulk of the 2D gas, and the voltmeter measures a seeming ``vanishing'' resistance. (In the bulk nothing essential happens: electrons just rotate around the cyclotron orbits responding to microwaves according to the classical equations of motion). The imaginary nonlocal conductivity $\sigma''_{xx}(\omega,q)$ oscillates as a function of $\omega$ and $\omega_c$ being a sum of terms 
\be -
\frac{\omega-k\omega_c}{(\omega-k\omega_c)^2+\gamma^2}
\ee 
with different $k$ ($\gamma=1/\tau$), 
so that the measured voltage $U_{xx}$ turns out to be close to what is shown in Figure \ref{fig:geometry}b, for details see Ref. \cite{Mikhailov11a} and Sections \ref{subsec:mirozrs} and \ref{sec:nonlin} below. 

The ponderomotive force is one of nonlinear electrodynamic effects which depends on the squared ac electric field and which may be observed in transport aa one sees from the MIRO experiments. Another nonlinear effect (which is the simplest and the most easily observed one) is the Joule heating: if the sample resistance $R(T)$ depends on the temperature of the electron gas, it varies under the action of the electromagnetic radiation, $T=T_0+\Delta T$ with $\Delta T\propto \sigma'(\omega)E^2(\bm r,t)$. Then the photoresistance $\Delta R=(\p R/\p T)\Delta T$ is proportional to the real part of the dynamic conductivity $\sigma'(\omega)$ and has a Lorenz-type resonance 
\be \propto 
\frac{1}{(\omega-\omega_c)^2+\gamma^2}
\ee 
at the cyclotron frequency (in an infinite sample), or at the magnetoplasmon frequency (if a sample has finite dimensions; see discussion of this issue in Ref. \cite{Mikhailov04a}). It is this effect that was observed in the early photoresistance experiment of Ref. \cite{Vasiliadou93}. 

The ponderomotive force effect is more difficult to observe. It is proportional to $\sigma''(\omega)\bm\nabla E^2(\bm r,t)$ and requires (a) ultraclean samples ($F_{pm}\propto \sigma''(\omega)$) and (b) a strongly inhomogeneous electric field ($F_{pm}\propto \nabla E^2$). In the MIZRS experiments these conditions were satisfied due to (a) the very high electron mobility [$\mu\simeq(15-30)\times 10^6$ cm$^2$/Vs], (b) the high microwave power
($\gtrsim 1$ mW/cm$^2$), and (c) the presence of metallic elements in the vicinity of the electron gas which led to the high concentration of the electric field and to sufficiently large gradients $\nabla E^2$. Notice that these metallic elements need not be real contacts touching the 2D gas; these can be, e.g., coplanar waveguides \cite{Andreev08,Andreev11} or other metallic structures placed sufficiently close to the 2D electron gas plane \cite{Bykov10b}, see discussion in Section \ref{subsubsec:contactless}.  

The rest of this paper consists of two essentially different parts. Section \ref{sec:miro} is a continuation of our previous publication \cite{Mikhailov11a} on the MIRO/ZRS effects. It contains a critique of the theories of Dmitriev, I\~narrea and Chepelianskii et al., as well as further details and explanations of the theory of Ref. \cite{Mikhailov11a} (after its publication we have got a number of questions and critical comments which will be answered and clarified here). Section \ref{sec:nonlin} reports a new result. We predict, on the basis of the same theoretical analysis that was done for MIRO \cite{Mikhailov11a}, that apart from the resistance oscillations, the system should demonstrate another nonlinear effect -- the second harmonic generation (SHG). This new effect should be experimentally observed under the same conditions as MIRO, and its experimental discovery could open a new applied-physics research direction -- the studies of nonlinear electrodynamic effects at microwave/terahertz frequencies (frequency multiplication, frequency mixing, etc.) in semiconductor high-electron-mobility 2D electron systems. In the last Section \ref{sec:conclusion} the results are summarized and  conclusions are drawn.

\section{MIRO and MIZRS: overview and comparison of different theories\label{sec:miro}}

In this Section different approaches to the theory of the MIRO/ZRS phenomena are discussed. 

\subsection{Main experimental features\label{subsec:prop}}

First, let us list the most important features of the MIRO/ZRS effects which have to be explained by a theory claiming that these effects are well understood.

\subsubsection{Ultra-clean samples, very high electron mobility\label{subsubsec:mobil}}

 One of the most important features of the discussed phenomena is that they have been observed in samples with an extremely hign electron mobility. Before 2000, when the typical mobility of GaAs quantum-well samples was of order  or below $1\times 10^6$ cm$^2$/Vs, there have been no indications on the MIRO of the type shown in Figure \ref{fig:geometry}b. In 2001 Zudov et al. \cite{Zudov01} and Ye et al. \cite{Ye01} reported, for the first time, on the microwave induced oscillations of the magnetoresistance. These oscillations were observed in samples with $\mu\simeq 3\times 10^6$ cm$^2$/Vs and were not yet so strong to form ``zero resistance'' intervals in the $R_{xx}(B)$ dependence. Then, in December 2002 and January 2003 Mani et al. \cite{Mani02} and Zudov et al. \cite{Zudov03} demonstrated $R_{xx}$-oscillations which clearly showed the ``vanishing resistance''. The electron mobility in their samples was $15\times 10^6$ cm$^2$/Vs and $25\times 10^6$ cm$^2$/Vs, respectively. Actually, only the publication of ``zero-resistance'' curves provoked great interest to the discussed phenomena. Thus the first characteristic feature of the MIRO/ZRS phenomena which is important for their understanding is that they are observed is ultraclean samples with the mobility exceeding $(10-15)\times 10^6$ cm$^2$/Vs. 

Sometimes it is stated that there exist exclusions from this rule. In 2008--2011 Wiedmann et al. published a series of papers in which resistance oscillations \cite{Wiedmann08} and zero resistance states \cite{Wiedmann10} were observed under the microwave irradiation in samples with the electron mobility $\lesssim 2\times 10^6$ cm$^2$/Vs. In these papers, however, one dealt with wide quantum wells with two occupied subbands \cite{Wiedmann10}, double \cite{Wiedmann08}, or triple \cite{Wiedmann09} quantum well samples with a high electron concentration [$\gtrsim (8-9)\times 10^{11}$ cm$^{-2}$]. In such systems the observed vanishing of $R_{xx}$ is actually a different effect. In these systems more than one electron subbands are typically occupied and the observed oscillations are closely related to the inter-subband transitions. For example, in \cite{Wiedmann10} the resistance lower than the dark one is observed at the same values of the $B$-field where the inter-subband maximum of $R_{xx}^0$ is seen in the absence of microwaves. 

In addition, in the Wiedmann's papers the microwave power was much higher than in the ultra-clean samples of Refs. \cite{Mani02,Zudov03}. In Ref. \cite{Wiedmann10} the ac electric field at which the ZRS becomes visible is estimated to be $\sim 4.2$ V/cm which corresponds to the power density $\sim 47$ mW/cm$^2$, while in the Mani paper \cite{Mani02} the estimated power level was below 0.75 mW/cm$^2$. 
Another exclusion was published by Bykov et al. in Ref. \cite{Bykov06}, where the ZRS were observed in samples with the mobility below $10^6$ cm$^2$/Vs, but also at the power level about five times higher than in Refs. \cite{Mani02,Zudov03}. 

One can therefore conclude that the original Mani-Zudov effect substantially depends on the sample quality and manifests itself better and better in samples with a growing mobility. It is also indicative that all really {\em new} features have been discovered in samples with extremely high electron mobility. For example, the giant spike on the double cyclotron frequency  $\omega=2\omega_c$ was observed in samples with the mobility of $30\times 10^6$ cm$^2$/Vs, Refs. \cite{Dai10,Dai11}.

\subsubsection{Nonlocal response\label{subsubsec:nonloc}} 

The second important feature of the MIRO/ZRS effects is that the $R_{xx}(B)$ oscillations are observed at the cyclotron harmonics $\omega=k\omega_c$ with a very large number $k$ (up to $k\simeq 10$). This is a very unusual behavior of the ac response. It is known that, when a 2D electron gas in a magnetic field is irradiated by electromagnetic waves, the cyclotron resonance at $\omega=\omega_c$ is observed. No resonances are seen at $\omega=2\omega_c$, $3\omega_c$, because the inter-Landau-level transitions between the levels $N$ and $N'$ are strictly forbidden for all $N\neq N'\pm 1$. 
This selection rule is violated only if the ac electric field acting on the electrons is strongly inhomogeneous, with the inhomogeneity scale of order of the cyclotron radius $r_c$. If a finite-size 2D-electron-gas sample is irradiated by microwaves, the typical inhomogeneity scale is of order of the sample dimensions or the wavelength of radiation. In the discussed experiments both these lengths exceed $100-200$ $\mu$m, while the cyclotron radius is of order of $0.5-10$ $\mu$m. The observation of many cyclotron harmonics can therefore be explained neither by the bulk  nor the near-edge effects (near the boundary of the 2D gas with the dielectric). Only near the contacts or, more generally, near sharp metallic edges can the electric field be inhomogeneous at the 1 $\mu$m scale, see Refs. \cite{Mikhailov06a,Mikhailov11a}. 

\subsubsection{Circular polarization\label{subsubsec:circular}}

A very important characteristic of the MIRO/ZRS effect is their independence of the circular polarization sense. The MIRO/ZRS effects were observed in samples with finite dimensions. That at microwave frequencies ($f\simeq 100$ GHz, wavelength 3 mm) the samples of the width $\sim 0.2$ mm cannot be considered as infinite was clear from the very beginning of the MIRO/ZRS story: in the microwave photoresistance experiment by Vasiliadou \cite{Vasiliadou93} in 1993, 
the resonance was observed not at the cyclotron, 
but at the magnetoplasmon frequency $\sqrt{\omega_c^2+\omega_p^2}$, where $\omega_p^2\propto 1/W$ is the 2D plasmon frequency and $W$ is the sample width, see discussion of this point in Ref. \cite{Mikhailov04a}. In principle, it is not a priori clear, whether the MIRO/ZRS effects have the bulk origin or is related to the boundaries of the sample. In an infinite sample in a magnetic field the natural basis is circular, i.e. the system response to the right- and left-circularly polarized radiation should be substantially different. The experiment \cite{Smet05} showed completely identical microwave induced $R_{xx}(B)$ oscillations for both circular polarizations. 

\subsubsection{Inelastic scattering and radiative decay\label{subsubsec:raddecay}} 

In view of the ``inelastic'' mechanism of the Dmitriev's theory one should discuss the role of the inelastic scattering in the MIRO/ZRS experiments. When the 2D electrons are excited by microwaves they continuously get energy from the ac electric field. In the stationary state they should continuously lose the same amount of energy. The scattering of electrons by impurities and (acoustic) phonons is quasi-elastic, i.e. the momentum relaxation time $\tau$ is much shorter than the energy relaxation time $\tau_{in}$, 
\be 
\gamma_{inelastic}^{impurities}=1/\tau_{in}\ll \gamma=1/\tau,
\ee
and the probability of the energy-loss processes due to the scattering is very low.  

In the ultra-clean 2D electron systems excited by microwaves there exists another, very efficient, mechanism of energy losses. This is the electron-{\em photon} scattering, or the radiative decay, i.e. electrons lose their energy by re-emitting electromagnetic waves. The efficiency of this process is characterized by the radiative decay rate 
\be 
\Gamma=\frac{2\pi n_s e^2}{mc};\label{raddecay}
\ee
the derivation of Eq. (\ref{raddecay}) and the corresponding discussion can be found in Ref. \cite{Mikhailov04a} and is not reproduced here. Comparing (\ref{raddecay}) with the (elastic) momentum relaxation rate $\gamma$, 
\be 
\frac\Gamma\gamma=\frac{2\pi n_se\mu}{c},
\ee
one sees that the ratio $\Gamma/\gamma$ depends on the electron mobility and equals one, $\Gamma/\gamma=1$, at $\mu\simeq 1.1\times 10^5$ cm$^2$/Vs for a typical electron density $n_s\simeq 3\times 10^{11}$ cm$^{-2}$. That is, in the MIRO/ZRS experiments 
\be 
\gamma_{inelastic}^{impurities}\ll \gamma\ll\Gamma\equiv\gamma_{inelastic}^{radiative} \label{inelastic-compar}
\ee
-- the inelastic scattering by impurities is many order of magnitude less probable process than the inelastic electron-photon scattering (the radiative decay).

\subsubsection{Influence of the parallel magnetic field\label{subsubsec:parfield}} 

In some experiments \cite{Yang06} it was shown that the MIRO oscillations are suppressed by a parallel magnetic field $B_\parallel$ of order of 1 T. This is also an unexpected result since the influence of $B_\parallel$ is usually associated with the spin effects. At so small magnetic field, $B_\parallel\simeq 1$ T, however, the spin effects are expected to be negligibly small. 

\subsubsection{Negative resistance\label{subsubsec:negres}} 

In most MIZRS experiments the measured resistance $R_{xx}$ drops down to zero in finite intervals of the magnetic field left from the points $\omega_c/\omega=1/k$. In Ref. \cite{Willett04}, however, a {\em negative} resistance (negative voltage $U_{xx}$) was observed. This effect is quite substantial: in Fig. 2 of Ref. \cite{Willett04} the absolute value of the ``negative'' $R_{xx}$ is about 50\% of $R_{xx}(B=0)$ at zero magnetic field.

\subsubsection{Zero conductance in Corbino geometry \label{subsubsec:Corbino}}

Apart from the zero resistance states in the Hall-bar geometry, a zero conductance effect was observed in some experiments \cite{Yang03,Bykov10b}. The vanishing conductance is observed in the same intervals of the magnetic field where the resistance zeros are seen in the Hall-bar geometry.

\subsubsection{Giant double cyclotron-frequency spike\label{subsubsec:spike}}

A new unexpected feature in the magnetic-field dependence $R_{xx}(B)$ was recently observed under microwave irradiation in Ref. \cite{Dai10}: a giant peak of the magnetoresistance at $B$ corresponding to the condition $\omega\simeq 2\omega_c$. This peak is very strong (about two orders of magnitude larger than the dark magnetoresistance at the same magnetic field) and narrow and was observed in a sample with an even higher mobility $\mu\simeq 30\times 10^6$ cm$^2$/Vs. This effect was then also studied in Refs. \cite{Dai11,Hatke11a,Hatke11b}.

\subsection{Inelastic-mechanism theory of Dmitriev et al.: Critical comments \label{DmitrievCritique}}

Although the theory of Dmitriev et al. \cite{Dmitriev03,Dmitriev04,Dmitriev05} was criticized \cite{Mikhailov04a} from the moment of its publication, the authors, as well as Zudov \cite{ZudovEP2DS}, continue to state that the MIRO/ZRS experiments are ``well understood'' within this theory. Let us consider how this theory explains the main experimental facts listed in Section \ref{subsec:prop}. 

\begin{figure}
\includegraphics[width=8.5cm]{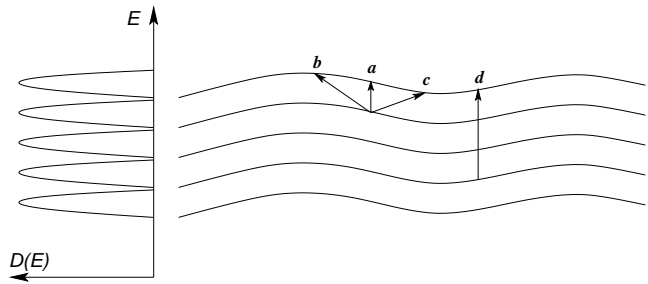}
\caption{\label{fig:landau} A schematic view of Landau levels and the corresponding density of states $D(E)$ in a sample with smooth disorder. The local density of states in an ultra-clean sample has the form of a set of delta-functions $\sum_n\delta(E-E_n)$; only the cumulative density of states averaged over the sample area is broadened. Out of transitions $\bm{a-d}$ only the vertical transitions $\bm a$ with $\omega=\omega_c$ are allowed. The indirect transitions $\bm b$ and $\bm c$ with $\omega>\omega_c$ and $\omega<\omega_c$, as well as the $\bm d$ transition with $\omega=k\omega_c$, $k>1$ are forbidden.  }
\end{figure}

The inelastic-mechanism theory of Dmitriev et al. \cite{Dmitriev03,Dmitriev04,Dmitriev05,Dmitriev12} suggests that the MIRO/ZRS effect has a bulk origin and is related to inelastic scattering of electrons within Landau levels substantially broadened due to a smooth disorder. Already these basic assumptions of the theory cause many questions. The effect is observed in samples with a very high electron mobility which corresponds to the mean free path of order of 100 $\mu$m. The cyclotron radius of electrons under the typical experimental conditions is about $1-2$ $\mu$m. This means that  most of electrons rotate around their cyclotron orbits very far from any impurity. The corresponding {\em local} density of states is then just a sum of delta-functions, $D_{loc}(E)\propto \sum_n\delta(E-E_n)$, where $E_n$ are the Landau-level energies. Of course, in the considered systems a smooth potential may be present (this is assumed in the Dmitriev's theory), which means that in the cumulative density of states the Landau levels are broadened indeed, Figure \ref{fig:landau}. However, for the optical transitions in the microwave field only the local density of states is relevant, i.e. only the transitions $\bm a$ with $\omega=\omega_c$ in Figure \ref{fig:landau} are allowed; the transitions $\bm b$ with $\omega>\omega_c$ and the transitions $\bm c$ with $\omega<\omega_c$ are indirect and therefore forbidden. The model \cite{Dmitriev03,Dmitriev04,Dmitriev05,Dmitriev12} does not thus correspond to the experimental reality. 

Also forbidden are the transitions $\bm d$ in Figure \ref{fig:landau} with $\omega=k\omega_c$ and $k\neq 1$. In the experiment the MIRO oscillations are seen at $k$ up to $k\simeq 10$. It is therefore unclear how the authors of \cite{Dmitriev03,Dmitriev04,Dmitriev05} could get from the identically vanishing inter-Landau-level matrix elements large $R_{xx}$ oscillations with very slowly varying amplitudes at large $k$. 

Third, the Dmitriev's theory predicts a strong difference of the MIRO response to the left- and right-circularly polarized microwave field. The experiment \cite{Smet05} clearly showed, in contrast, that MIRO oscillations are identical for both types of the circular polarization. In Ref. \cite{Dmitriev12} the authors try to save their theory speculating on the influence of the radiative decay (\ref{raddecay}) and on the transformation of the circular polarization to the linear one near the contacts which was first discussed in our paper \cite{Mikhailov06a} and then finally used for the development of a complete theory of MIRO/ZRS in \cite{Mikhailov11a}. But, first, an elementary calculation shows that taking into account the radiative decay does not eliminate the difference between the system response to the left and right circular polarization. Second, it is clear that the metal contacts screen the external field only in the very vicinity of the contacts and not in the bulk of the 2D system\cite{Mikhailov06a,Mikhailov11a}. The latter fact is evidently confirmed by the same experiment of Smet \cite{Smet05} which demonstrated a very large difference in the absorption spectra (a bulk effect) for two circular polarizations and no difference in the MIRO oscillations (thus, evidently, \textit{not} a bulk effect). Thus, the Dmitriev's theory also contradicts the most crucial experimental fact (Section \ref{subsubsec:circular}) -- the  insensitivity of the MIRO/ZRS effects to the circular polarization sense \cite{Smet05}.

Then, according to Dmitriev et al. \cite{Dmitriev03,Dmitriev04,Dmitriev05}, ``a distinctive feature'' of their theory ``is the inelastic electron scattering''. It is this feature that they used to argue against another bulk theory -- the so called ``displacement'' mechanism \cite{Durst03}. However, the probability of the inelastic scattering of electrons by impurities is {\em several orders of magnitude smaller} than the probability of the inelastic electron-photon scattering (Section \ref{subsubsec:raddecay}, as well as Ref. \cite{Mikhailov04a}). That is, the theory \cite{Dmitriev03,Dmitriev04,Dmitriev05} takes into account tiny negligible processes but ignores much more important in the ultra-clean samples inelastic radiative effects.

The Dmitriev's theory is also unable to explain the strong suppression of the MIRO oscillations in parallel magnetic fields \cite{Yang06} (Section \ref{subsubsec:parfield}). 

No bulk theory of MIZRS, including the Dmitriev's one, can explain the negative voltage $U_{xx}$ that was observed in Ref. \cite{Willett04} (Section \ref{subsubsec:negres}). If the potential difference $U_{xx}$ was indeed proportional to the bulk resistivity $\rho_{xx}$ of the electron gas, the negative resistance would mean the real instability of the system, with a possibility to extract energy from it, to cool the system by irradiation (the negative Joule heat) and other fantastic things. 

The $2\omega_c$-spike discovered  \cite{Dai10} in samples with the mobility of $30\times 10^6$ cm$^2$/Vs (Section \ref{subsubsec:spike}) also turned out to be a great surprise for the theory of Dmitriev et al. In order to explain it, they would have to argue that in a system with {\em even weaker} disorder, the {\em forbidden} inter-Landau-level $2\omega_c$-transition leads to a {\em huge} enhancement of the bulk resistance because of their {\em disorder-based} mechanism. 

All the above arguments against the theory \cite{Dmitriev03,Dmitriev04,Dmitriev05} are qualitative. Consider the quantitative aspects of the Dmitriev's theory. We refer to the ``central result'' of Ref. \cite{Dmitriev05}, the photoconductivity $\sigma_{ph}$ at arbitrarily strong values of both dc and ac electric fields \cite{Dmitriev05},
\be 
\frac{\sigma_{ph}}{\sigma_{dc}^D}=1+2\delta^2\left[1-\frac{{\cal P}_\omega \frac{2\pi\omega}{\omega_c}\sin \frac{2\pi\omega}{\omega_c}+4{\cal Q}_{dc}}{1+{\cal P}_\omega\sin^2 \frac{\pi\omega}{\omega_c} +{\cal Q}_{dc}}\right],\label{Dmcentres1}
\ee
where $\sigma_{dc}^D$ is the dc Drude conductivity, $\delta=\exp\left(-\pi/\omega_c\tau_q\right)$,  ${\cal P}_\omega$ is a parameter proportional to the power of the incident electromagnetic wave, 
\be 
{\cal P}_\omega=\frac{\tau_{in}}{\tau_{tr}}\left(\frac{eE_\omega v_F}{\hbar\omega}\right)^2 \frac{\omega_c^2+\omega^2}{(\omega^2-\omega_c^2)^2},
\label{DmP}
\ee
and ${\cal Q}_{dc}$ is proportional to the squared dc electric field (we will omit this term considering only the Ohmic regime typical for MIRO/ZRS experiments). In Eqs. (\ref{Dmcentres1}) and (\ref{DmP}) $\tau_{tr}$, $\tau_{in}$, and $\tau_q$ are the transport, inelastic and zero-$B$ single-particle relaxation times, and $v_F$ is the Fermi velocity. The longitudinal photoresistivity $\rho_{ph}$ can then be written in the form \cite{Dmitriev05}
\be 
\frac{\rho_{ph}}{\rho_{dc}^D}=
1+2e^{-a /X}\left[1-\frac{{\cal P}_\omega^{(0)}  \frac{1+X^2}{(1-X^2)^2} \frac{2\pi}{X}\sin \frac{2\pi}{X}} {1+{\cal P}_\omega^{(0)} \frac{1+X^2}{(1-X^2)^2}\sin^2 \frac{\pi}{X} }\right]\label{Dmcentres2}
\ee
where $X=\omega_c/\omega$ is the dimensionless magnetic field and 
\be 
a=\frac{2\pi}{\omega\tau_q}, \ \ \ 
{\cal P}_\omega^{(0)}=\frac{\tau_{in}}{\tau_{tr}}\left(\frac{eE_\omega v_F}{\hbar\omega^2}\right)^2 \label{params}
\ee
are $B$-independent parameters. 

\begin{figure}
\includegraphics[width=8.1cm]{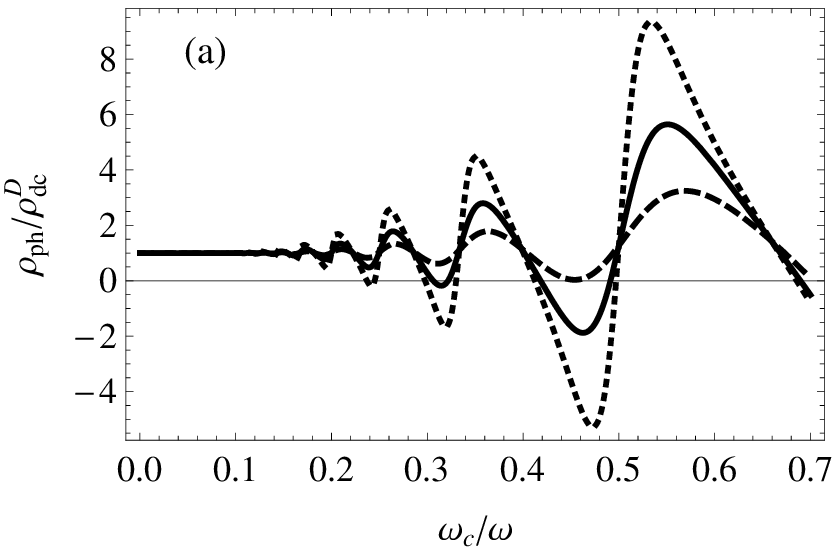}
\includegraphics[width=8.1cm]{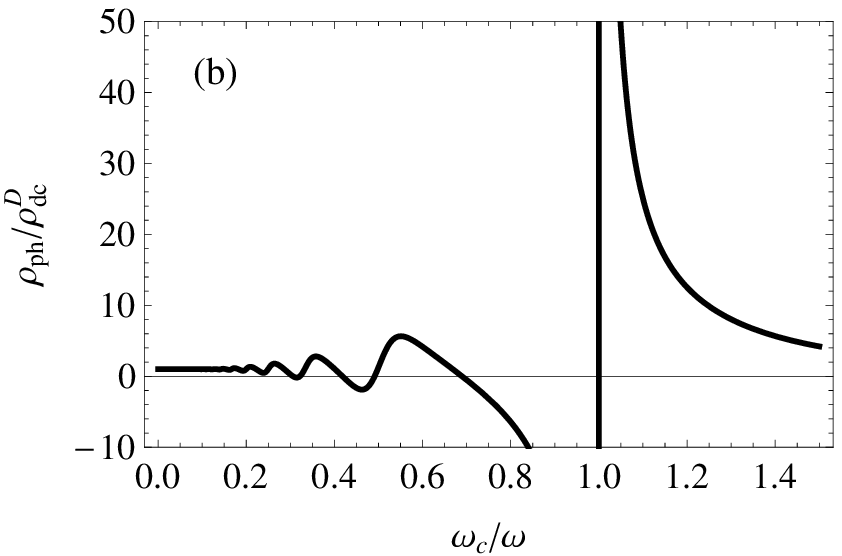}\\
\includegraphics[width=8.1cm]{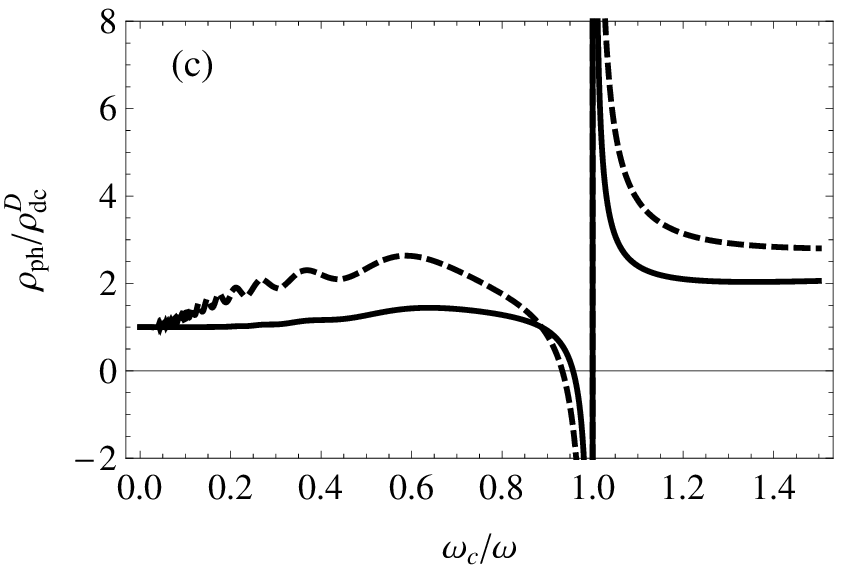}
\caption{\label{fig:Dmitriev} (a) The microwave photoresistivity (normalized to the dark Drude value) calculated in Ref. \cite{Dmitriev05} at $\omega\tau_q=2\pi$ and the power parameters ${\cal P}_\omega^{(0)}=0.24$, 0.8, and 2.4. These values of ${\cal P}_\omega^{(0)}$ are about three orders of magnitude larger than in the experiments, see discussion in the text. (b) The same dependence (only the curve for ${\cal P}_\omega^{(0)}=0.8$ is shown) in a broader range of the magnetic field, covering the fundamental cyclotron harmonic. The $k=1$ harmonic is huge as compared to the $k=2,3,\dots$ harmonics, in a strong disagreement with experiments. (c) The microwave photoresistivity at the experimentally realistic power parameter ${\cal P}_\omega^{(0)}=0.005$. The solid and dashed curves correspond to $\omega\tau_q=2\pi$ and $\omega\tau_q=10\pi$ respectively (see discussion in the text). Being properly used (with realistic experimental parameters) the theory \cite{Dmitriev05} does not demonstrate any agreement with experiments. }
\end{figure}

Figure \ref{fig:Dmitriev}(a) shows the photoresistivity (\ref{Dmcentres2}) as a function of the $B$-field at $a=1$ and three different values of the power parameter, ${\cal P}_\omega^{(0)}=0.24$, 0.8, and 2.4. We emphasize that this Figure is plotted with the help of the analytical result (\ref{Dmcentres1}) derived in Ref. \cite{Dmitriev05} and reproduces Fig. 2 from Ref. \cite{Dmitriev05}. The authors claim that these results describe the MIRO experiments. However, the curves in Figure \ref{fig:Dmitriev} (it is identical to Fig. 2 from Ref. \cite{Dmitriev05}) and the formula (\ref{Dmcentres2}) causes at least two questions. 

First, why the calculated oscillations are shown only at $\omega_c/\omega\lesssim 0.7$? Where is the main resistance oscillation at $\omega_c/\omega\simeq 1$? Using the Dmitriev's formula (\ref{Dmcentres2}) one can plot the calculated in Ref. \cite{Dmitriev05} photoresistance in a broader range of $\omega_c/\omega$; the result is shown in Figure \ref{fig:Dmitriev}(b) for ${\cal P}_\omega^{(0)}=0.8$. The agreement with the experiment is no longer so good: the first-harmonic feature ($k=1$) is \textit{much larger} than the other ones ($k>1$) which \textit{disagrees} with experiments. Actually, at $\omega_c/\omega\simeq 1$ the calculated photoresistance diverges, see formula (\ref{Dmcentres2}), although all relaxation times ($\tau_{tr}$, $\tau_q$, $\tau_{in}$) are taken into account in the theory. 

Now the second question. What is the power of radiation which corresponds to Figure 2 in Ref. \cite{Dmitriev05} (Figure \ref{fig:Dmitriev}(a) here)? In Ref. \cite{Dmitriev05} the authors carefully studied  different relaxation times and came to a conclusion that their realistic values under the conditions of MIRO experiments are: $\tau_{in}\simeq\tau_{tr}\approx 1$ ns and $\tau_q\simeq \tau_{tr}/100\approx 10$ ps. We agree with these estimates. The value $a=1$ used in Figure \ref{fig:Dmitriev}(a) follows from the last number for $\tau_q$ at the typical microwave frequency $f=100$ GHz of the MIRO experiments (see e.g. Ref. \cite{Mani02}). Now, what about the power parameter ${\cal P}_\omega^{(0)}$? Substituting the experimental data of Mani et al. \cite{Mani02} into Eq.  (\ref{params}) ($f\approx 100$ GHz, the power density $\simeq 1$ mW/135 mm$^2$, the density of electrons $n_s=3\times 10^{11}$ cm$^{-2}$), we get 
\be 
{\cal P}_\omega^{(0)}=4.65\times 10^{-3}.
\ee 
That is, the theoretically calculated $\rho_{ph}$-oscillations shown in Fig. 2 of Ref. \cite{Dmitriev05} and reproduced here in Figure \ref{fig:Dmitriev}(a) were obtained at \textit{three orders of magnitude larger microwave power than in the experiments}. Let us plot the Dmitriev's photoresistance (\ref{Dmcentres2}) \textit{at the experimentally realistic power level} of ${\cal P}_\omega^{(0)}=5\times 10^{-3}$. The result is shown by the black solid curve in Figure \ref{fig:Dmitriev}(c). The perfect agreement with experiment completely disappeared. The authors understand the weakness of their theory in this point (but do not plot experimentally relevant curves!) and speculate in Ref. \cite[Section IX]{Dmitriev05} that, perhaps, the $\tau_q$ time is in fact longer. We plot the Dmitriev's formula (\ref{Dmcentres2}) at the five times bigger $\tau_q$. The result shown by the dashed curve in Figure \ref{fig:Dmitriev}(c) is not much better. 

During the last ten years Dmitriev et al. (as well as many experimentalists, e.g. \cite{ZudovEP2DS}) claimed that the MIRO/ZRS phenomena are ``well understood'' within their theory. One sees that this statement is enormously optimistic.

\subsection{I\~narrea's work}

A very large number of theoretical papers (over 20) was produced by I\~narrea after his first publication in Physical Review Letters \cite{Inarrea05}. The formula for the photoconductivity $\sigma_{xx}$ of the 2D electron gas derived there   [Eq. (5) in \cite{Inarrea05}] shows that the conductivity $\sigma_{xx}$ (not conductance!) tends to infinity in the Ohmic regime $E_{dc}\to 0$, depends on the area of the sample $S$ and tends to infinity when $S$ grows, $\sigma_{xx}\propto S/E_{dc}$ (here $E_{dc}$ is the dc electric field). The Hall conductivity $\sigma_{xy}$ in the same Phys. Rev. Lett. publication does not depend on the electron density but is proportional to the density of impurities. Since all subsequent I\~narrea's publications are based on this first paper \cite{Inarrea05} and these misprints have never been corrected, we do not discuss this theory further.

\subsection{Chepelianskii's theory}

Chepelianskii and Shepelyansky \cite{Chepelianskii09} tried to explain the MIRO oscillations analyzing the classical dynamics of 2D electrons near the edge of the 2D gas (i.e. near its boundary with a dielectric) under the action of microwaves and in the presence of scatterers. This attempt is partly reasonable since it employs the classical approach (relevant under the experimental conditions), and considers not a bulk but an edge mechanism. But the theory \cite{Chepelianskii09} cannot explain all experimentally observed features, in particular, the microwave induced conductance oscillations in the Corbino geometry \cite{Yang03}: in the Corbino rings there is no boundary 2D gas -- dielectric, but the effect is present. In addition, the value of the electric field parameter $\epsilon=eE_{ac}/\omega p_F$, at which the calculated resistance oscillations become sufficiently large ($\epsilon\simeq 0.06$, see Fig. 3 of Ref. \cite{Chepelianskii09}), is about 30 times larger than the experimental values ($\epsilon\simeq 0.002$ for parameters of Ref. \cite{Mani02}). The observation of MIZRS corresponding to the mechanism of Ref. \cite{Chepelianskii09} would therefore require \textit{three orders of magnitude stronger microwave power} than it was in the real experiments.

\subsection{The ponderomotive-forces theory \cite{Mikhailov11a}\label{subsec:mirozrs}}

A coherent theory of the MIRO/ZRS effects which explained all experimentally observed features was developed in 2011 in our paper \cite{Mikhailov11a}. In this Section we give a brief overview of this theory and answer all questions and criticisms which were caused by that publication. 

\subsubsection{Overview of the theory  \cite{Mikhailov11a}\label{subsubsec:overview_my_theory}}

As seen from Figure \ref{fig:geometry}, the shape of the investigated Hall-bar sample reminds a spider with a small semiconductor ``body'' and several thin long metallic ``legs'' -- contact wires. In conventional magnetoresistance measurements (without microwaves) the contact wires do not play any active role. Under the microwave irradiation, however, they serve as antennas, focusing the microwave radiation in small near-contact areas, like lightning rods. The near-contact ac electric field $E_c$ is strongly inhomogeneous and substantially exceeds the incident-wave field $E_0$, $E_c\gg E_0$, see Ref. \cite{Mikhailov06a}. For example, near the sharp edge of a two-dimensional conductor occupying the half-plane $z=0$, $x<0$, the electric field $E_x$ is known to be proportional to $E_x\propto x^{-1/2}$, Ref. \cite[problem 3 to \S 3]{Landau8}. In addition, this field is linearly polarized, independent of the circular polarization of the incident wave \cite{Mikhailov06a}, since near metallic surfaces the field is always normal to the metal boundary. This linearly polarized, strong and strongly inhomogeneous electric field $\bm E(\bm r,t)$ acts on the near-contact electrons by a time-independent second-order ponderomotive force $\bm F_{pm}(\bm r)\propto \bm\nabla\langle E^2(\bm r,t)\rangle_t$, where $\langle \dots \rangle_t$ means averaging over time. In the above example $E_x^2\propto 1/x$, and $F_{pm}\propto 1/x^{2}$. The force $\bm F_{pm}(\bm r)$ is proportional to the imaginary part of the nonlocal conductivity of the 2D electron gas $\sigma''(\omega,\omega_c)$, and, therefore, changes its direction depending on the sign of $\omega-k\omega_c$: it repels electrons from the contacts at $\omega_c\lesssim\omega/k$ and attracts them to the contacts at $\omega_c\gtrsim\omega/k$, thus forming depletion or accumulation regions in the near-contact areas. Since the near-contact ac electric field is strongly inhomogeneous on the $r_c$-scale, the effect is seen not only around the fundamental frequency $\omega\simeq \omega_c$, but also around higher harmonics $\omega\simeq k\omega_c$. In the depletion regime, $\omega_c\lesssim\omega/k$, the ponderomotive potential may become larger than the Fermi energy near the contacts. In this case the strong ponderomotive force isolates the bulk of the 2D gas from the contacts, thus leading to the vanishing voltage $U_{xx}$ (which is no longer proportional to the bulk magnetoresistance $R_{xx}$), and to the seeming ``zero-resistance'' states. The near-contact density of the 2D electrons is then proportional to the Boltzmann factor $\sim\exp(-P/T)$, where $P$ is the microwave power, thus explaining the activation dependence of the resistance minima, $\ln R_{xx}^{min} \propto -P/T$, Ref. \cite{Willett04}. The ponderomotive force is a collisionless effect; it does not need any disorder and plays more and more important role when the sample quality is improved, i.e. when the electron mobility increases. This explains why the MIRO/ZRS phenomena have been observed only in samples with an extremely high electron mobility.

For the quantitative description of the above outlined physics the reader is referred to Ref. \cite{Mikhailov11a} (see also Section \ref{sec:nonlin} below).
Now we discuss some further details of the theory in the form of ``Questions and Answers'' (Q \& A). 

\subsubsection{Q \& A: ``Depletion layers should modify the two-terminal resistance''\label{subsubsec:two-terminalresistance}}

\begin{figure}
\includegraphics[width=8.5cm]{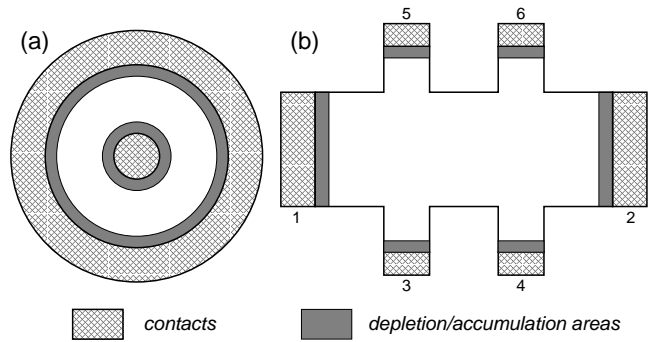}
\caption{\label{CorbHallgeom} The geometry of (a) the Corbino disk and (b) the Hall-bar sample under the intense microwave irradiation. The gray areas near the contacts show the microwave induced depletion/accumulation regions.}
\end{figure}

Consider the near-contact processes in more detail. The formation of the near-contact depletion regions at $\omega_c\lesssim\omega/k$ immediately explains the conductance minima in the Corbino-geometry experiment, Figure \ref{CorbHallgeom}, see Ref. \cite{Mikhailov11a}. In the Hall-bar geometry, however, a question on the two-contact resistance arises (the question of Ramesh Mani and Alexei Chepelianskii). Indeed, the depletion regions are formed not only near the side (potential) contacts 3 -- 6 but also near the current contact 1 -- 2 (Figure \ref{CorbHallgeom}). Therefore, one can think that the formation of the depletion regions near the contacts 1 -- 2 will increase the two-terminal, and hence the Hall resistance measured between the contacts 3 -- 5 (or 4 -- 6), which would then contradict the experiment (microwaves do not influence the Hall resistance). 

This point was discussed in Ref. \cite[Section II D and Fig. 5 there]{Mikhailov11a} but a more direct and convincing answer has been recently given by Alexei Chepelianskii \cite{Chepelianskii13privcomm}. Using the finite elements technique, he numerically calculated the distribution of the electric potential and the current in a Hall bar of the type shown in Figure \ref{CorbHallgeom}, with and without the near-contact depletion regions. He found that the two-terminal resistance varies only in low magnetic fields, $\omega_c\tau\lesssim 1$. In the regime of strong $B$, $\omega_c\tau\gg 1$, the depletion regions do not change the two-terminal (and Hall) resistance. Since in the MIRO/ZRS experiments the $\omega_c\tau$-parameter often exceeds 100, this result perfectly agrees with experiments.

\subsubsection{Q \& A: ``If MIRO is an edge effect, its amplitude should depend on geometry''}

Another question (criticism) to the ponderomotive near-contact-mechanism theory \cite{Mikhailov11a} was formulated by Dmitriev et al. in Ref. \cite{Dmitriev12}: ``\dots experimentally, no dependence of the MIRO amplitude on the sample dimension or geometry, characteristic of the edge effects, has been reported so far''. It is not completely clear to me why the MIRO amplitude should depend on the sample dimension or geometry in the near-contact mechanism of MIRO, but in order to clarify this point we consider here the MIRO and ZRS formation in some more detail.

\begin{figure}
\includegraphics[width=8.5cm]{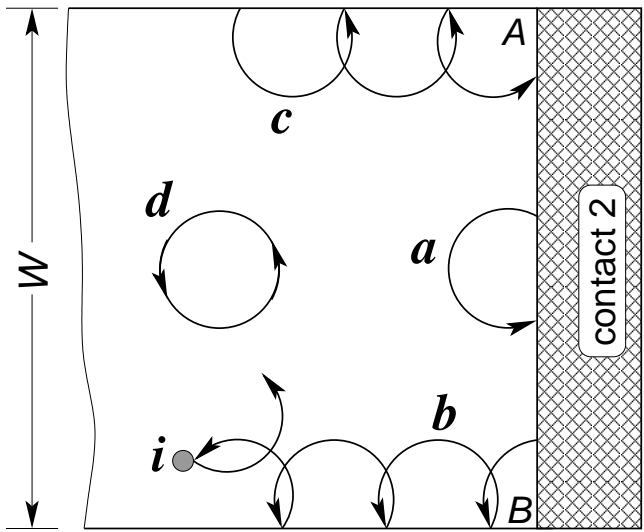}
\caption{\label{fig:contactsA} Electron trajectories in the vicinity of the right contact in the presence of the magnetic field $B$. Since the mean free path is much larger than the cyclotron radius, electrons $\bm a$ which leave the contact in points far from the corners $A$ and $B$ experience the Lorentz force and return back to the contact. Electrons $\bm b$ which start from the points close to the corner $B$ start to skip along the lower sample boundary; electrons $\bm c$ arrive from the left contact skipping along the upper boundary and enter the contact near the corner $A$. Bulk electrons $\bm d$ rotate around the cyclotron orbits inside the sample. Edge (skipping) electrons may be scattered to the bulk (and visa versa) by impurities $\bm i$.  }
\end{figure}

Consider the motion of electrons in the vicinity of the right contact in an ultra-clean sample, Figure \ref{fig:contactsA}. Let, first, no dc current flow in the sample and no microwaves irradiate it. Electrons of the 2D gas and those of the metallic contact should be in the thermodynamic equilibrium, i.e. the number of electrons which leave the contact should be equal to the number of electrons which enter the contact. Let the magnetic field $B$ be about $0.5$ T -- a typical value in the MIRO/ZRS experiments. Then the cyclotron radius $r_c$ is about $0.5 -$few $\mu$m which is much smaller than the sample width ($\gtrsim 200$ $\mu$m) and the mean-free-path $l_{mfp}$ of the electrons ($\simeq 100$ $\mu$m). Electrons $\bm a$ (Figure \ref{fig:contactsA}), which leave the contact with a certain velocity somewhere far from the corners $A$ and $B$, have a very good chance to be returned back to the contact by the Lorentz force ($r_c\ll l_{mfp}$). Only electrons $\bm b$, which start their motion from the points sufficiently close to the corner $B$ (at the distance $\sim r_c$), leave the contact and begin to skip along the lower boundary 2D gas -- dielectric to the left with a rather high average velocity (of order of the Fermi velocity $v_F$). Similarly, electrons $\bm c$ skip along the upper boundary of the 2D gas to the right and enter the contact near the corner $A$. In equilibrium the number of electrons which enter the corner $A$ equals the number of electrons which leave the corner $B$. Apart from electrons skipping along the lower and upper boundaries with the velocity $\sim v_F$ there are electrons $\bm d$ which just rotate around the cyclotron orbits in the bulk of the 2D gas. The edge (skipping) electrons may be scattered to the bulk by an impurity $\bm i$ and visa versa. In equilibrium the probabilities of these processes are equal.  

Let now a dc bias be applied between the right and left contacts, so that the chemical potential of the right contact $\mu_R$ is somewhat higher than the chemical potential of the left contact $\mu_L$ (which is kept grounded, $\mu_L=0$). Then the number of $\bm b$-electrons is bigger than the number of $\bm c$-electrons, i.e. the lower (upper) boundary gets negatively (positively) charged. As a result, the Hall field $E_H\sim (\mu_R-\mu_L)/eW$ appears, and the bulk electrons $\bm d$ also start to move with the drift velocity $v_{dr}\sim cE_H/B\ll v_F$. The probabilities of the edge-bulk scattering also change: the scattering processes ``lower edge $\to$ bulk'' and `` bulk $\to$ upper edge'' have larger probabilities than the opposite processes. 

\begin{figure}
\includegraphics[width=8.5cm]{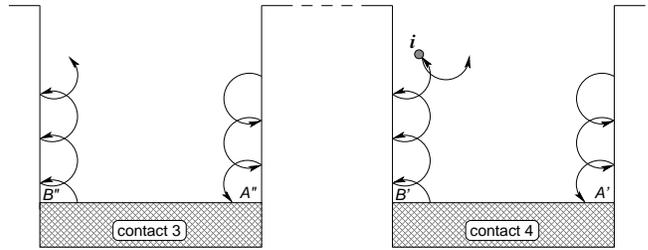}
\caption{\label{fig:contactsB} Electron trajectories near the side contacts 3 and 4. No current flows into the sample through these contacts, therefore the number of electrons entering the contact 4 (contact 3) at the corner $A'$ ($A''$) equals the number of electrons leaving the contact 4 (contact 3) at the corner $B'$ ($B''$). In equilibrium the contacts 3 and 4 have the same potential, therefore the number of electrons leaving the contact 4 at the corner $B'$ equals the number of electrons entering the contact 3 at the corner $A''$. If a net current flows in the sample, $\mu_R>\mu_L$, the scattering processes ``lower edge $\to$ bulk'' are more probable than the opposite processes, and not all electrons from the corner $B'$ arrive at the corner $A''$. As a result a potential difference $U_{xx}=\mu_4-\mu_3>0$ is established. }
\end{figure}

Now consider the side contacts 3 and 4, Figure \ref{fig:contactsB}. Electrons skipping along the lower boundary enter the contact 4 at the corner $A'$. Since the potential of this contact should remain constant, the same amount of electrons should leave the contact 4 at the corner $B'$. The same should be valid at the potential contact 3: the number of electrons which enter this contact at the corner $A''$ should be equal to the amount of electrons which leave it at the corner $B''$. If no dc bias is applied between the contacts 1 and 2, the number of electrons which leave the contact 4 at the corner $B'$ equals the number of electrons which enter the corner $A''$ of the contact 3; as a result, the contacts 3 and 4 are at the same potential, $\mu_3=\mu_4=0$. 

If the net current flows in the sample ($\mu_R> \mu_L$) the number of electrons which leave the contact 4 is not equal to the number of electrons which enters the contact 3. A part of them is lost on the way between the side contacts due to the scattering to the bulk, Figure \ref{fig:contactsB}. As a result the chemical potential $\mu_4$ of the contact 4 turns out to be larger than the chemical potential $\mu_3$ of the contact 3, $\mu_4>\mu_3>\mu_L=0$. A potential difference $U_{xx}=\mu_4-\mu_3>0$ is established. In the ultra-clean samples the density of impurities is very low, therefore $U_{xx}$ is small as compared to the Hall, or two-contact voltage. 

\begin{figure}
\includegraphics[width=8.5cm]{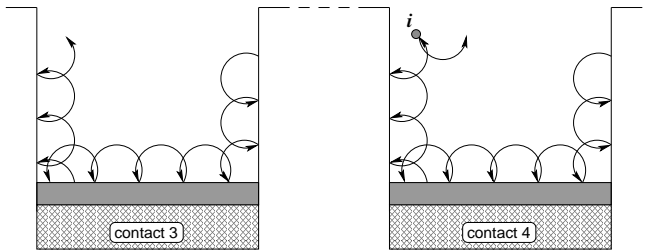}
\caption{\label{fig:contactsC} The same as in Figure \ref{fig:contactsB} but under the microwave irradiation in the ZRS regime, i.e. when the ponderomotive force repels electrons from the contacts. The gray areas show the microwave induced depletion regions. Electrons repelled by the ponderomotive potential partly pass by the contacts which leads to a reduction of the measured potential difference $U_{xx}$.}
\end{figure}

Now, if the sample is irradiated by microwaves and the $B$-field corresponds to the condition of the ``vanishing resistance'', i.e. the ponderomotive force repel electrons from the contacts, depletion layers are formed near the contacts, Figure \ref{fig:contactsC}. Only a certain portion of electrons (say, $p$ \%) skipping along the boundary of the 2D gas can overcome the ponderomotive potential barrier and enter the contacts; all other electrons pass by as shown in Figure \ref{fig:contactsC}. As a result, the chemical potentials $\mu_4^{mw}$ and $\mu_3^{mw}$ of the contacts are reduced by a factor $p$ as compared to their dark values, $\mu_3^{mw}\sim p \mu_3$, $\mu_4^{mw}\sim p\mu_4$. The potential difference $U_{xx}^{mw}$ is then also by a factor $p$ smaller than in the dark, $U_{xx}^{mw}=\mu_4^{mw}-\mu_3^{mw}\sim p(\mu_4-\mu_3)=pU_{xx}$, which is perceived as the reduction of the sample resistance. The reduction coefficient  $p$ is related to the density factor ${\cal N}=n_c/n_s$ defined in \cite{Mikhailov11a} (the ratio of the density of electrons $n_c$ at the contact to their density in the bulk of the sample). 

Now, returning back to the criticism of Ref. \cite{Dmitriev12} we see that the sample-dimension or geometry dependence is not characteristic of the ponderomotive-force effect. Therefore, the fact that ``\dots experimentally, no dependence of the MIRO amplitude on the sample dimension or geometry \dots has been reported so far'' is in perfect agreement with our theory. 

\subsubsection{Q \& A: How to explain the negative bias $U_{xx}$?}

In the previous Section we assumed that the contacts have identical properties, i.e. the factor $p$ is the same for both contacts 3 and 4. If, however, the side contacts have (accidentally) different properties, so that, for instance, the same microwave power induces different ponderomotive potentials at the contacts 3 and 4, the $p$-factors may be different, $p_3\neq p_4$. Then the measured potential difference $U_{xx}^{mw}=\mu_4^{mw}-\mu_3^{mw}\sim p_4\mu_4-p_3\mu_3$ may have both signs. This gives a simple explanation of the negative-$U_{xx}$ experiment \cite{Willett04} discussed in Section \ref{subsubsec:negres}. 

\subsubsection{Q \& A: How to explain the influence of the parallel magnetic field?}

This point has been already discussed in our previous paper \cite[Section II E 4 there]{Mikhailov11a}; here we briefly mention this point for completeness. 

The height of the pondermotive potential depends on the screening properties of the metallic contacts. If the contacts are ideal (have an infinite conductivity $\sigma_m\to\infty$) and infinitely thin in the $z$-direction normal to the plane of the 2D electron gas then the electric field and the ponderomotive force diverge near the contact as $1/\sqrt{x}$ and $1/x^2$ respectively, see Section \ref{subsubsec:overview_my_theory}. Real contacts have a finite thickness in the $z$-direction and their dielectric properties quite substantially depend on the $B$-field of order of $1-1.5$ T (the sum of $B_\parallel$ and $B_\perp$ which is relevant for the three-dimensional contact). The parallel magnetic field thus influences not the electronic properties of the 2D electron gas (this would require much stronger $B_\parallel$), but modifies the screening properties of the metallic contacts. It reduces the amplitude of the ponderomotive potential and thus leads to the suppression of MIZRS.

\subsubsection{Q \& A: What about contactless experiments?\label{subsubsec:contactless}}

\begin{figure}
\includegraphics[width=8.5cm]{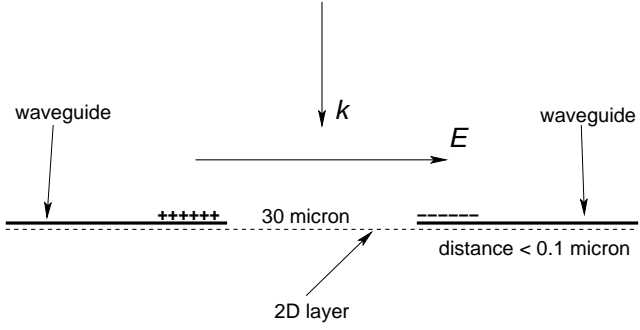}
\caption{\label{AndreevExper} The geometry of the Andreev's experiment \cite{Andreev11}. The 2D layer is covered by a metallic coplanar waveguide with the width of the open area (the distance between metals) $\sim 30$ $\mu$m. The distance between the 2D gas and the waveguide is smaller than 0.1 $\mu$m. The incident electromagnetic wave with the wave vector $\bm k$ induces oscillating charges at the edges of the waveguide and hence a strong and strongly inhomogeneous electric field, Figure \ref{fieldingap}. This produces a ponderomotive potential and modifies the 2D electron density in exactly the same way as was discussed in our Refs. \cite{Mikhailov06a,Mikhailov11a}. The true contact between the metallic waveguide and the 2D gas in not needed. }
\end{figure}

There exist several experiments in which MIRO were observed in systems without contacts (a comment of Stefan Wiedmann). Indeed Andreev et al. \cite{Andreev08,Andreev11} developed a contactless technique to measure the conductivity $\sigma_{xx}$ of a 2D electron gas irradiated by microwaves. They covered a 2D quantum-well system by a metallic coplanar waveguide, Figure \ref{AndreevExper}, irradiated it by a high-frequency electromagnetic wave ($f=118$ GHz) and studied the propagation of a low-frequency signal ($f_{low}=400$ MHz) along the waveguide. The coplanar waveguide transmission is determined by the real part of the diagonal conductivity $\sigma_{xx}'$ of the 2D gas, therefore measuring the transmission one could determine $\sigma_{xx}'$. The real contacts to the 2D gas were placed far from the propagation channel of the probe ($f_{low}$) wave, so that their influence on the results of the measurements was excluded. The experiment \cite{Andreev11} showed microwave induced conductivity oscillations. 

\begin{figure}
\includegraphics[width=8.5cm]{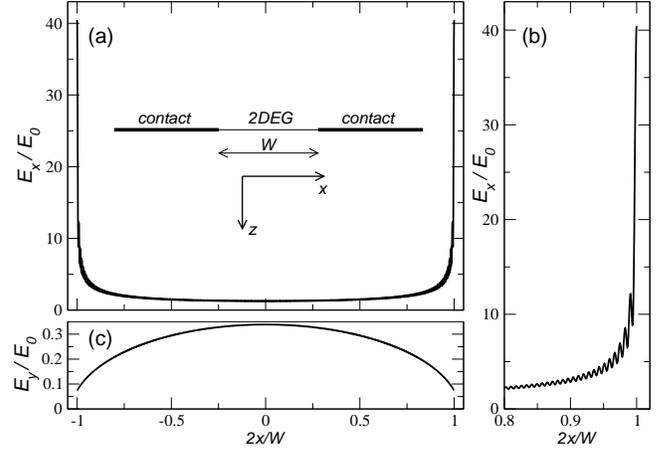}
\caption{\label{fieldingap} The calculated electric field inside the gap between the contact wings, in the absence of the 2D electrons. This picture is taken from Ref. \cite{Mikhailov11a}, see this paper for details of calculations and the values of parameters. (a), (b) Electric field is perpendicular to the boundary 2D electron gas -- contact; (c) electric field is parallel to the boundary 2D electron gas -- contact. Inset in (a) shows the geometry of the 2D stripe between the two contact wings.}
\end{figure}

This experiment does not, however, contradict the ponderomotive-force theory of the MIRO/ZRS effects. As was shown in Ref. \cite{Mikhailov11a} and the above discussion (Section \ref{subsubsec:overview_my_theory}) the strong ponderomotive force appears in the system near the sharp edges of metallic contacts. In the experiment \cite{Andreev11} the real contacts were absent, but a metallic waveguide was placed on top of the structure in the vicinity of the 2D gas. Figure \ref{fieldingap} taken from \cite{Mikhailov11a} shows the result of my calculation of the electric field in a gap between two conducting half-planes lying in the plane $z=0$ at the distance $W$ from each other. The inset shows the geometry of the problem; notice that in this calculation we assumed that there are no 2D electrons in the gap between the two metallic half-planes, i.e. the calculated field is the external field for the 2D electrons. The calculation shows, in agreement with the textbook results \cite[problem 3 to \S 3]{Landau8}, that the field $E_x(x,z=0)$ diverges near the metallic edges $x=\pm W/2$. Now, it is clear that at a small distance $d$ from the plane $z=0$  the electric field $E_x(x,z=d)$ will also be much stronger than the field of the incident electromagnetic wave and strongly inhomogeneous, if the distance $d$ is much smaller than the distance $W$ between the metallic half-planes. In the experiment \cite{Andreev11} the distance $d$ between the waveguide and the 2D electron gas in the $z$-direction was smaller than 0.1 $\mu$m, while the width $W$ between the metallic half-planes was equal to 30 $\mu$m. At so small ratio $d/W\lesssim 1/300$ all the physics described in Ref. \cite{Mikhailov11a} and in Section \ref{subsubsec:overview_my_theory} remains valid: the incident powerful electromagnetic radiation induces a strong ponderomotive potential near the waveguide edges, this leads to the depletion of the 2D gas under the waveguide edges, which then modifies the propagation of the probe low-frequency wave along the waveguide. 

Let us quantitatively estimate, whether the incident microwave power in the experiment \cite{Andreev11} was sufficient to deplete the 2D gas under the waveguide edges. The density of electrons in Ref. \cite{Andreev11} was about $2\times 10^{11}$ cm$^{-2}$ which corresponds to the Fermi energy of $\sim 6.4$ meV. According to \cite{Mikhailov11a}, the ponderomotive potential near the main frequency harmonic $\omega=\omega_c$ is
\be 
U_{pm}(x)=
\frac{e^2E_x^2(x)}{8m\omega_c}
\Bigg(
\frac{ \omega-\omega_c}{(\omega-\omega_c)^2+\gamma^2}
-
\frac{ \omega+\omega_c}{(\omega+\omega_c)^2+\gamma^2}
\Bigg).
\ee 
The maximum of this potential, as a function of $B$, lies at $|\omega-\omega_c|\sim \gamma$ and, as a function of $x$, at the point $x\simeq d$, i.e. 
\be 
U_{pm}^{max}\simeq 
\frac{e^2E_{max}^2}{16 m\omega_c
\gamma}\sim \frac{e^2\tau E_{max}^2}{16 m\omega}
\sim \frac{e\mu E_{max}^2}{32\pi f},
\ee 
where $E_{max}\simeq E_x(x\simeq d)$. According to Figure \ref{fieldingap} the maximum field at the distance $\simeq W/300$ from the metallic edge of the waveguide is at least 50 times larger than the field of the incident wave. The incident power density in \cite{Andreev11} was about 10 mW/cm$^2$ which corresponds to the electric field in the incident wave of order of 2 V/cm. Assuming that the field $E_{max}$ is about 50 times bigger we get $E_{max}\sim 100$ V/cm. At the mobility $\mu\sim 10^7$ cm$^2$/Vs and the frequency $f\sim 100$ GHz we then obtain that the depth of the ponderomotive potential well,
\be 
U_{pm}^{max}
\sim \frac{e\mu E_{max}^2}{32\pi f}\sim 10\textrm{ meV},
\ee 
is larger than the Fermi energy without microwaves which perfectly confirms our interpretation. 

Another contactless experiment was performed by Bykov et al. in Ref. \cite{Bykov10b}. The conductance of a 2D electron system was measured in the Corbino geometry by a capacitive method by placing ring-shaped electrodes on the planar surface of the samples. In this experiment the distance $d$ between the 2D gas and the sample surface was equal 105 nm, and the distance between the inner and outer Corbino electrodes was $\sim 1.5$ mm. The ratio $d/W=0.7\times 10^{-5}$ was two orders of magnitude smaller than in the Andreev's experiment \cite{Andreev11}, so that the observation of very clear ``zero-conductance'' states in Ref. \cite{Bykov10b} perfectly confirms the ponderomotive-force theory \cite{Mikhailov11a}. 

\subsubsection{Q \& A:  1/4-phase of resistance oscillations}

Michel Dyakonov noticed \cite{DyakonovMIRO} that in the ponderomotive-force theory \cite{Mikhailov11a} the minima/maxima of the $R_{xx}$ oscillations do not lie at $\omega/\omega_c=k\pm 1/4$ as it is the case in the Dmitriev's theory and was observed in experiments, but are related to the scattering rate parameter $\gamma/\omega_c$. 

\begin{figure}
\includegraphics[width=8.5cm]{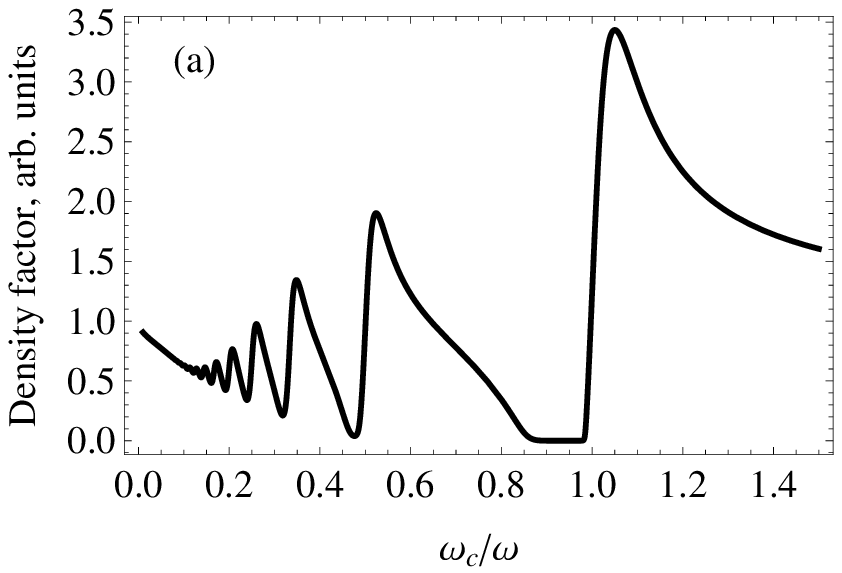}\\
\includegraphics[width=8.5cm]{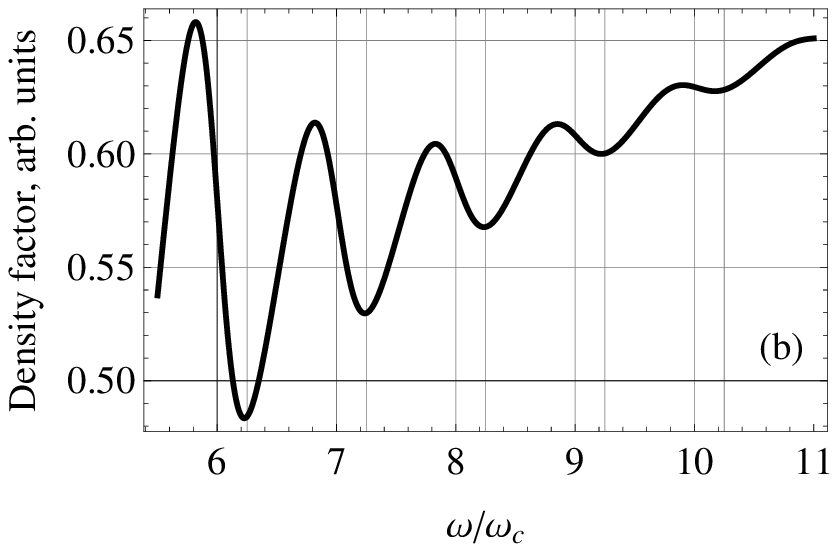}
\caption{\label{fig:dyak} The density factor ${\cal N}$ as a function of (a) $\omega_c/\omega$ and (b) $\omega/\omega_c$ at $\gamma/\omega=0.05$, $v_F/\omega l=2$, $T/E_F=0.08$, $E_c/E_0=20$, and the power parameter ${\cal P}=0.005$ (for designations see Ref. \cite{Mikhailov11a}). The vertical thin lines in (b) correspond to $\omega/\omega_c=k$ and $k+1/4$. One sees that at large $k$ the minima of the ${\cal N}$ curve are very close to the $k+1/4$ points. }
\end{figure}

The answer to this comment is as follows. The calculated in \cite{Mikhailov11a} density factor ${\cal N}$, which determines the measured value of $U_{xx}$, depends on many input parameters (see discussion in Section II E in Ref. \cite{Mikhailov11a}), some of which are not well known. For example, the most important but unknown ingredient of the theory -- the contacts -- were modeled by a very simple expression for the near-contact electric field $E(x)=E_c/\sqrt{1+x/l}$, which not necessarily describes the real situation correctly. But even within these simplified assumptions and approximations it is not difficult to reproduce the 1/4-phase with the theory \cite{Mikhailov11a}, see Figure \ref{fig:dyak}, choosing a set of parameters different from those used in Ref. \cite{Mikhailov11a}. Figure \ref{fig:dyak} shows that at $k\gtrsim 6$ the minima and maxima lie indeed at $\omega/\omega_c=k\pm \phi$ with $\phi\approx 1/4$. 

At smaller values of $k$ the phase $\phi$ decreases and gets closer to the scattering factor $\gamma/\omega_c$. However, this deviation does not contradict the experiments. The value of the phase $\phi$ was actually a subject of disputes. For example, Mani et al. \cite{Mani02,Mani04} claimed that the 1/4-phase rule is valid for all magnetic fields, while Zudov \cite[Figure 1 of this paper]{Zudov04} argued that the phase $\phi$ tends to $\pm 1/4$ only at large values of $k$ ($k\gtrsim 6$), while at smaller $k$ it noticeably decreases (our results of Figure \ref{fig:dyak} well reproduce this behavior of the phase). Moreover, in their first papers \cite{Mani02,Zudov03} Mani and Zudov actually made opposite statements: according to \cite{Mani02}, $\delta R_{xx}$ vanishes at integer values of $\omega/\omega_c$, while Zudov \cite{Zudov03} stated that $\delta R_{xx}$ has maxima at these values. This uncertainty was clarified later (Ref. \cite{Zudov04}) but even the fact, that there was such an uncertainty, indicates that the small-$k$ maxima lie very close to $\omega/\omega_c=k$, especially in the very high mobility samples with $\gamma/\omega\ll 1$. It may also be noticed that the misinterpretation in the early Zudov experiment \cite{Zudov03} could be due to the higher mobility of his samples ($\mu=25\times 10^6$ cm$^2$/Vs) as compared to the samples of Ref. \cite{Mani02} ($\mu=15\times 10^6$ cm$^2$/Vs).

Thus, the $k$-dependence of the phase $\phi$ predicted by the ponderomotive-force theory is in a very good agreement with experimental data: it gives $\phi\simeq \pi/4$ at large $k$ ($\gtrsim 5-6$) and smaller values at $k\simeq 1$. 

As for the Dmitriev's theory which supposedly demonstrates the $1/4$-phase rule, we have seen in Section \ref{DmitrievCritique} that, being properly used (with realistic experimental parameters) this theory does not demonstrate any agreement with experiments, see Figure \ref{fig:Dmitriev}(c).

\subsubsection{Q \& A: Linear polarization experiments by Mani}

\begin{figure}
\includegraphics[width=8.5cm]{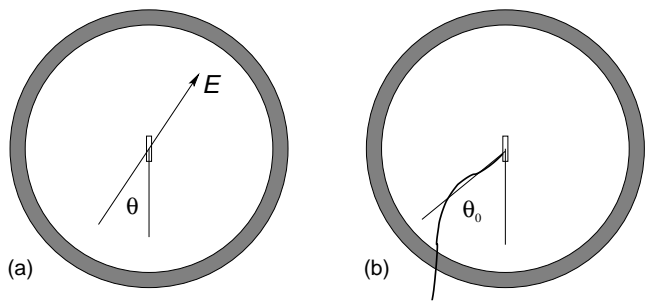}
\caption{\label{fig:maniwaveguide} (a) The geometry of the experiments \cite{Mani11,Ramanayaka12}. A small-size Hall-bar sample is placed inside a large-diameter circular waveguide. The linear polarization angle $\theta$ between the electric field vector $\bm E$ and the Hall-bar axis is varied in this experiment. (b) A possible interpretation of the sinusoidal dependence of the MIRO amplitudes (\ref{manitheta}) with the shift angle $\theta_0$: if the metallic wires used for the $R_{xx}$-measurements are oriented at an angle $\theta_0$ with respect to the Hall-bar axis, one can expect a maximum of $R_{xx}$ amplitude at $\theta=\theta_0$.   }
\end{figure}

One more question which may arise to the ponderomotive-force theory is how to interpret recent results of the Mani group \cite{Mani11,Ramanayaka12} on the linear-polarization sensitivity of the MIRO oscillations. In these papers the authors irradiated a small (0.4-mm-wide) Hall-bar sample by  linearly polarized microwaves propagating in a circular waveguide with the 11-mm internal diameter, Figure \ref{fig:maniwaveguide}(a). The electric field vector of the incident wave was rotated during the experiment and the amplitude of the $R_{xx}$ oscillations was measured as a function of the angle $\theta$, see Figure \ref{fig:maniwaveguide}(a). It was found that $R_{xx}(\theta)$ has a sinusoidal $\theta$ dependence 
\be 
R_{xx}(\theta)=A+C\cos^2(\theta-\theta_0)
\label{manitheta}
\ee 
with a maximum at the angle $\theta_0$ which varied between $47^\circ$ and $83^\circ$ dependent on the type of the MIRO maxima or minima and the direction of the magnetic field. 

Commenting on the results of Refs. \cite{Mani11,Ramanayaka12} we would like to notice that microwave experiments with metallic elements in the irradiated area should always be performed and interpreted with great care. Being irradiated by microwaves the metallic wires serve as antennas re-radiating the waves and substantially distorting the initial power distribution and the polarization of the electromagnetic wave in the waveguide. The preferential polarization direction at $\theta=\theta_0$ in these experiments may be simply related to the orientation of the lead wires, Figure \ref{fig:maniwaveguide}(b) (then the near-contact electric field producing the ponderomotive potential will be somewhat stronger), and not to any physical processes inside the sample.

\subsubsection{Q \& A: What about the $2\omega_c$ spike?}

The origin of the recently observed \cite{Dai10,Dai11,Hatke11a,Hatke11b}, in the vicinity of the 2nd cyclotron harmonic $\omega=2\omega_c$, giant photoresistivity spike is not yet understood. It is noticeable, however, that the first theoretical attempt to explain this effect \cite{Volkov13preprint} employs the ideas of the ponderomotive-force theory \cite{Mikhailov06a,Mikhailov11a}. In this paper it is assumed that a strongly inhomogeneous near-contact electric field causes a \textit{parametric resonance} in the system, which then leads to the observed giant $U_{xx}$-spike. Since the strongest parametric resonance is known \cite{Landau1} to be observed at the double resonance frequency (i.e. at $\omega=2\omega_c$ for the fundamental cyclotron resonance in the MIRO experiments) and under the action of an inhomogeneous external force \cite{Mikhailov98a}, the idea of Ref. \cite{Volkov13preprint} sounds very reasonable. 

\subsection{MIRO and MIZRS: Conclusions}

The MIRO and ZRS phenomena are being experimentally investigated already more than ten years in many groups in the world. A very large number of different, often unexpected, features have been discovered. All these characteristic features can be understood within the ponderomotive-force theory \cite{Mikhailov11a}. It consistently explains the microwave frequency, polarization, power, magnetic field, mobility, and temperature dependencies of the MIRO and ZRS effects. 

The ponderomotive-force phenomenon is a non-trivial nonlinear electrodynamic effect. The observation of such nonlinear phenomena is possible only in sufficiently strong electromagnetic fields and in systems where the scattering and disorder effects are negligibly small. Therefore the discovery of the MIRO and ZRS turned out to be possible only in ultra-clean GaAs quantum-well samples with an extremely high electron mobility. 

Such a collisionless solid-state electron plasma may become a gold-mine for discoveries of other nonlinear electrodynamic phenomena. In the next Section we predict one of such effects. We show that in the same ultra-clean GaAs quantum-well samples another nonlinear  effect should be observed: a second harmonic generation. This effect, directly related to the MIRO/ZRS phenomena, can be used for creating microwave and terahertz sources of radiation and is therefore potentially of great practical importance.

\section{Second harmonic generation \label{sec:nonlin}}

The formulas for the SHG effect can be derived in the same way as it was done in Ref. \cite{Mikhailov11a} for MIRO/ZRS phenomena. 

\subsection{Equations of motion}

In the previous Sections we saw that an electromagnetic wave, incident on the boundary ``2D electron gas -- a sharp metallic contact'', creates a strong and strongly inhomogeneous ac electric field near this boundary, Figure \ref{fieldingap}. Consider the (almost) collisionless motion of a 2D electron in the close vicinity of such a contact. We assume that the 2D gas and the contact layer lie in the plane $z=0$, and the 2D gas (the contact) occupies the half-plane $x>0$ ($x<0$). Further, we assume that the external permanent magnetic field points in $z$-direction, $\bm B=(0,0,B)$, and the microwave 
electric field near the contact has the form 
\be 
E_x(x,t)=E_x(x)\cos\omega t.\label{inhom-field}
\ee 
In zeroth order in $E_x$ an electron rotates around the cyclotron orbits
\begin{equation} 
\left(
\begin{array}{r}
x^{(0)} (t) \\
y^{(0)} (t)\\
\end{array}
\right)=
\left(
\begin{array}{r}
x_0 +r_c \sin(\omega_ct+\phi_0) \\
y_0-r_c \cos(\omega_ct+\phi_0) \\
\end{array}
\right),\label{rsol}
\end{equation} 
where $x_0$, $y_0$, and $\phi_0$ are determined by initial conditions. Since the microwave field (\ref{inhom-field}) is inhomogeneous on the cyclotron radius scale, the electron experiences different forces in different parts of its trajectory. In the first order in $E_x$, the force 
\ba 
F_x^{(1)}(t)&=&-eE_x[ x^{(0)}(t)]\cos\omega t
\nonumber \\ &=&
-eE_x[ x_0 +r_c \sin(\omega_ct+\phi_0)]\cos\omega t
\ea 
contains an infinite number of harmonics with the frequencies $\pm \omega +k\omega_c$, 
\ba
F_x^{(1)}(t) &=& -\frac e2 \sum_{k=-\infty}^\infty \epsilon_k(x_0) e^{ik(\phi_0-\pi/2)}
\nonumber \\ &\times&
\left(e^{i(k\omega_c +\omega) t}+e^{i(k\omega_c-\omega) t}\right),
\label{force}
\ea
where the factors $\epsilon_k(x_0)$ are determined by the inverse Fourier transform of the field, s. Ref. \cite{Mikhailov11a},
\begin{equation} 
\epsilon_k(x_0)= \epsilon_{-k}(x_0)=\frac 1{\pi}\int_{0}^\pi E_x(x_0+r_c\cos\xi)\cos k\xi d\xi .\label{epsdef}
\end{equation} 
Solving equations of the electron motion, 
\begin{equation} 
{\bm {\dot r}=\bm v},\hspace{3mm}
m{\bm {\dot v}}=-\frac ec {\bm v}\times {\bm B} -\gamma m {\bm v}+F_x(t){\bm e}_x  ,\label{eqs}
\end{equation}
with the force $F_x(t)=F_x^{(1)}(t)$ from (\ref{force}), we get the first-order correction to the coordinate,
\ba
   x^{(1)}(t)&=& \frac e{2m }\sum_{k=-\infty}^\infty \epsilon_k(x_0) e^{ik(\phi_0-\pi/2)}
\nonumber \\ &\times&
\left(\frac{e^{i(k\omega_c +\omega) t}}{(k\omega_c +\omega-i\gamma)^2-\omega_c^2} \right.
\nonumber \\ &+&
\left.
\frac{e^{i(k\omega_c-\omega) t}}{(k\omega_c -\omega-i\gamma)^2-\omega_c^2}
\right)  .\label{x1(t)}
\ea
In the second order in $E_x$, the force 
\be
 F_x^{(2)}(t)=
-e \cos\omega t \frac{\p E_x[x^{(0)}(t)]}{\p x}x^{(1)}(t), \label{vdot5}
\ee 
acting on the electron, is then
\begin{eqnarray} 
F_x^{(2)}(t)
&=&
-\frac {e^2}{4m }  
\sum_{k,k'=-\infty}^\infty 
\frac{\partial \epsilon_{k'} (x_0)}{\partial x_0} 
\epsilon_k(x_0)
e^{i(k'+k)(\phi_0-\pi/2)}  \nonumber \\ &\times&
\left(
\frac{e^{i[(k+k')\omega_c +2\omega] t}+
e^{i(k+k')\omega_c  t}}{(k\omega_c +\omega-i\gamma)^2-\omega_c^2}
\right.
\nonumber \\ &+&
\left.
\frac{e^{i(k+k')\omega_c  t}+e^{i[(k+k')\omega_c-2\omega] t}}{(k\omega_c -\omega-i\gamma)^2-\omega_c^2}
\right) .\label{force2nd}
\end{eqnarray}  
It depends on the initial conditions of the electron motion, $x_0$ and $\phi_0$. Averaging Eq. (\ref{force2nd})  over $\phi_0$, we get
\ba
\langle F_x^{(2)}\rangle_\phi&=&
-\frac {e^2}{16m \omega_c} (1+e^{i2\omega t})
\nonumber \\ &\times&
\frac{\partial }{\partial x_0} 
\sum_{k=-\infty}^\infty 
\frac{\epsilon_{k+1}^2 (x_0)-\epsilon_{k-1}^2 (x_0)}{k\omega_c +\omega-i\gamma}
+c.c.,\label{force-average}
\ea
where $c.c.$ means the complex conjugate. 

\subsection{Static response: The ponderomotive force}

The force (\ref{force-average}) is proportional to the factor $(1+e^{i2\omega t})$ and thus contains two effects. The time-independent term is the ponderomotive force considered in Ref. \cite{Mikhailov11a} and in Section \ref{sec:miro} above. The corresponding ponderomotive potential,
\ba
U_{pm}(x_0)&=&
\frac {e^2}{8m \omega_c} 
\sum_{k=1}^\infty 
\Big(\epsilon_{k-1}^2 (x_0)-\epsilon_{k+1}^2 (x_0)\Big)
\nonumber \\ &\times&
\left(
\frac{\omega- k\omega_c }{(\omega-k\omega_c )^2+\gamma^2}
-\frac{\omega+k\omega_c }{(\omega+k\omega_c )^2+\gamma^2}
\right),\nonumber \\ \label{pond-pot}
\ea
depends on the distance $x_0$ of an electron from the contact. Since $E_x(x)$ and, hence, $\epsilon_k(x_0)$, dramatically grow approaching the contact, electrons closest to the contact (at $x_0\simeq r_c$) experience the largest force. The near-contact change of the density, which determines the visible change of the measured voltage $U_{xx}$, is then determined by the density factor \cite{Mikhailov11a},
\be 
{\cal N}=\frac{n_s(r_c)}{n_s}=\frac T{E_F}F\left(\frac{E_F-U_{pm}(r_c)}T\right),\label{densfact}
\ee
which is defined as the ratio of the near-contact density, $n_s(r_c)$, to the density of electrons $n_s$ far from the contacts, in the bulk of the 2D layer. The function  
\be 
F(z)=\int_0^\infty \frac {dx}{1+\exp(x-z)}
\ee
in (\ref{densfact}) is the Fermi integral. 
This formula was obtained in \cite{Mikhailov11a} and describes experimentally observed resistance oscillations and zero-resistance states under the influence of microwaves, for details see \cite{Mikhailov11a}. 

\subsection{Dynamic response: Second harmonic generation}

The second contribution to the force (\ref{force-average}) oscillates in time with the frequency $2\omega$. Substituting it in the equation of motion (\ref{eqs}) we calculate the second-order velocity 
\ba 
&& \left(
\begin{array}{l}
v_x^{(2)}(t) \\
v_y^{(2)}(t) \\
\end{array}
\right)
=
\frac {1}{2m } \frac{\partial U_{pm}(x_0)}{\partial x_0} 
\nonumber \\ &\times& 
\left(
\begin{array}{r}
-\sin(2\omega t) \left(\frac{2\omega-\omega_c}{(2\omega-\omega_c)^2+\gamma^2}
+ \frac {2\omega+\omega_c}{(2\omega+\omega_c)^2+\gamma^2}\right)\\
\cos (2\omega t) \left(\frac{2\omega-\omega_c}{(2\omega-\omega_c)^2+\gamma^2}
- \frac {2\omega+\omega_c}{(2\omega+\omega_c)^2+\gamma^2}\right)\\
\end{array}
\right)
.
\label{v2}
\ea
One sees that  electrons oscillate with the double microwave frequency $2\omega$, both in the $x$- and $y$-directions. The oscillation amplitude depends on their position in the sample and is maximal near the contacts. Averaging Eq. (\ref{v2}) over the sample width and multiplying the result by $-en_s$ we get the ac current at the frequency $2\omega$,
\ba 
&& \left(
\begin{array}{l}
 j_x^{(2)}  \\
 j_y^{(2)}  \\
\end{array}
\right)
=
 \frac {e n_s(U_{pm}^L-U_{pm}^R)}{2m W}
\nonumber \\ &\times&
\left(
\begin{array}{r}
-\sin(2\omega t) \left(\frac{2\omega-\omega_c}{(2\omega-\omega_c)^2+\gamma^2}
+\frac {2\omega+\omega_c}{(2\omega+\omega_c)^2+\gamma^2}\right)
\\
\cos (2\omega t)\left(\frac{2\omega-\omega_c}{(2\omega-\omega_c)^2+\gamma^2}
-\frac {2\omega+\omega_c}{(2\omega+\omega_c)^2+\gamma^2}\right)
 \\
\end{array}
\right).
\label{j2av}
\ea
The electric field of the radiated wave is then $E^{(2)}_{x,y}\simeq 2\pi j^{(2)}_{x,y}/c$, and the intensity of radiation polarized perpendicularly ($\perp$) and parallel ($\parallel$) to the boundary 2D gas -- contact is 
\be 
J_\perp^{(2)}=\frac {\pi}{c}(j_x^{(2)})^2,\ \ \ J_\parallel^{(2)}=\frac {\pi}{c}(j_y^{(2)})^2.\label{intens}
\ee
Using Eqs. (\ref{intens}), (\ref{j2av}) and (\ref{pond-pot}) one can calculate the $B$- and $\omega$-dependence of the emitted second-harmonic intensity and to estimate its absolute value. This is done below.

\begin{figure}
\includegraphics[width=8.5cm]{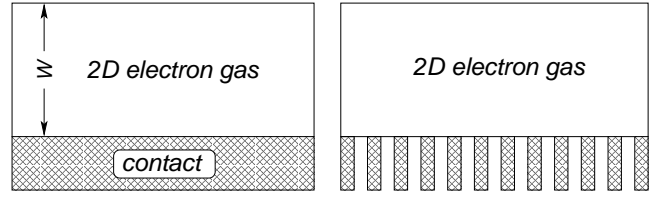}
\caption{\label{fig:emitelement} Possible realizations of the second harmonic emitter. The emitting element should be asymmetric (the contact on one side) and have a moderate width ($W\lesssim 20-50$ $\mu$m). In the right embodiment one could achieve a higher concentration of the microwave power near the tips of the contact wires and hence a larger ponderomotive potential. To increase the efficiency of the SHG-emitter an array of such emitting elements should be used with the total area exceeding the radiation wavelength.}
\end{figure}

A few comments should be made on the current formula (\ref{j2av}). The quantities $U_{pm}^L$ and $U_{pm}^R$ there are the amplitudes of the ponderomotive potential at the left and right side of the sample. In a symmetric structure, similar to that shown in the Inset to Fig. \ref{fieldingap}(a), the amplitudes $U_{pm}^L$ and $U_{pm}^R$ are equal, and the second-harmonic radiation vanishes. To avoid this the emitting elements should be asymmetric, for example, their left side borders on a metallic contact while the right side borders on a dielectric (the contacts 5 and 6 in Figure \ref{CorbHallgeom}(b) are absent). Then one of the amplitudes $U_{pm}^L$ or $U_{pm}^R$ equals zero (we will assume below that $U_{pm}^R=0$). The basic emitting elements may thus have the form shown in Figure \ref{fig:emitelement}. The width $W$ of the 2D-gas area should be rather small (several $r_c$, i.e. $\gtrsim 10-20$ $\mu$m) since the ac current amplitude in proportional to $1/W$, Eq. (\ref{j2av}). In order to increase the overall efficiency of the device the elements shown in Figure \ref{fig:emitelement} may be arranged in an array.

\begin{figure}
\includegraphics[width=8.5cm]{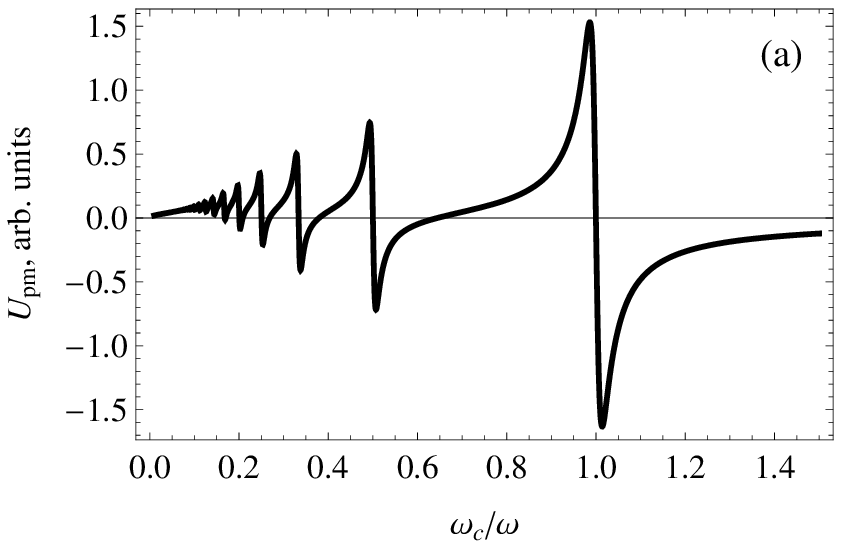}\hfill \includegraphics[width=8.5cm]{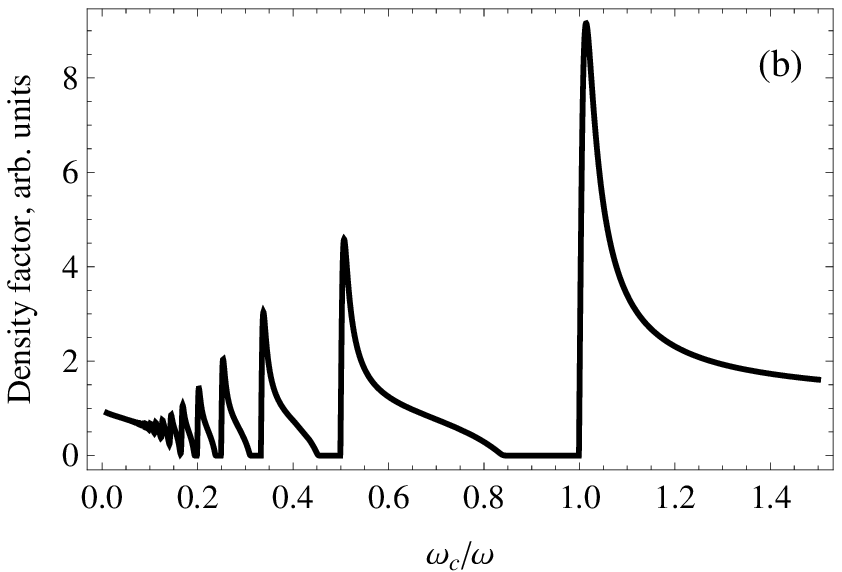}\\
\caption{\label{fig:zrs} (a) The ponderomotive potential (\ref{pond-pot}) and (b) the density factor (\ref{densfact}) as a function of $\omega_c/\omega$. The parameters used: $\omega/\gamma= 72$, $E_c/E_0=20$, $v_F/\omega l=2$, and $T/E_F=0.02$. For the inhomogeneous electric field amplitude we used the model (\ref{inhom-field-model}). Notice that the results do not essentially differ from those of Ref. \cite{Mikhailov11a} where another model for $E_x(x)$ was used.}
\end{figure}

\begin{figure}
\includegraphics[width=8.5cm]{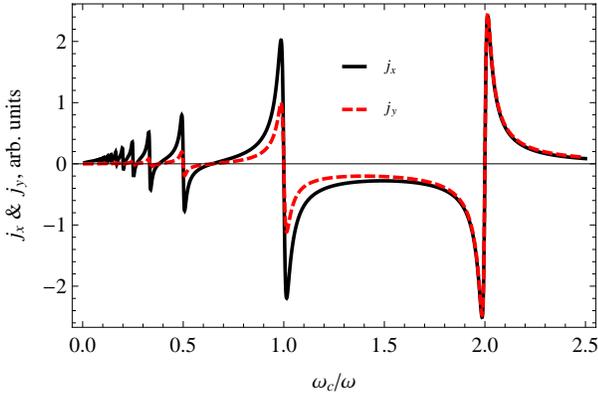}\hfill 
\caption{\label{fig:current} The amplitude of the second-harmonic current (\ref{j2av}) as a function of $\omega_c/\omega$. The parameters are the same as in Figure \ref{fig:zrs} (except $T/E_F$ which is not needed). In contrast to the ponderomotive potential (\ref{pond-pot}), Figure \ref{fig:zrs}(a) the sign of the curves does not matter since the current varies in time and the figure shows the current amplitude. Experimentally measured quantities are proportional to the squared amplitudes, Figure \ref{fig:intens}.}
\end{figure}

\begin{figure}
\includegraphics[width=0.48\columnwidth]{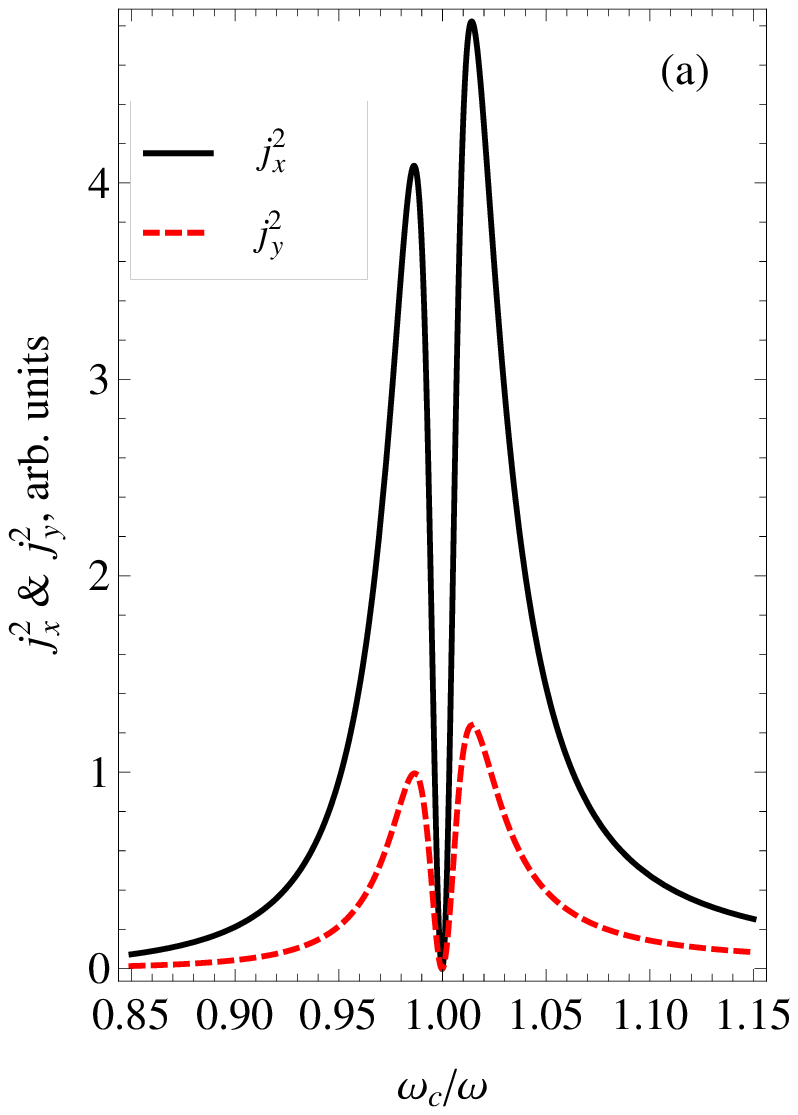}
\includegraphics[width=0.48\columnwidth]{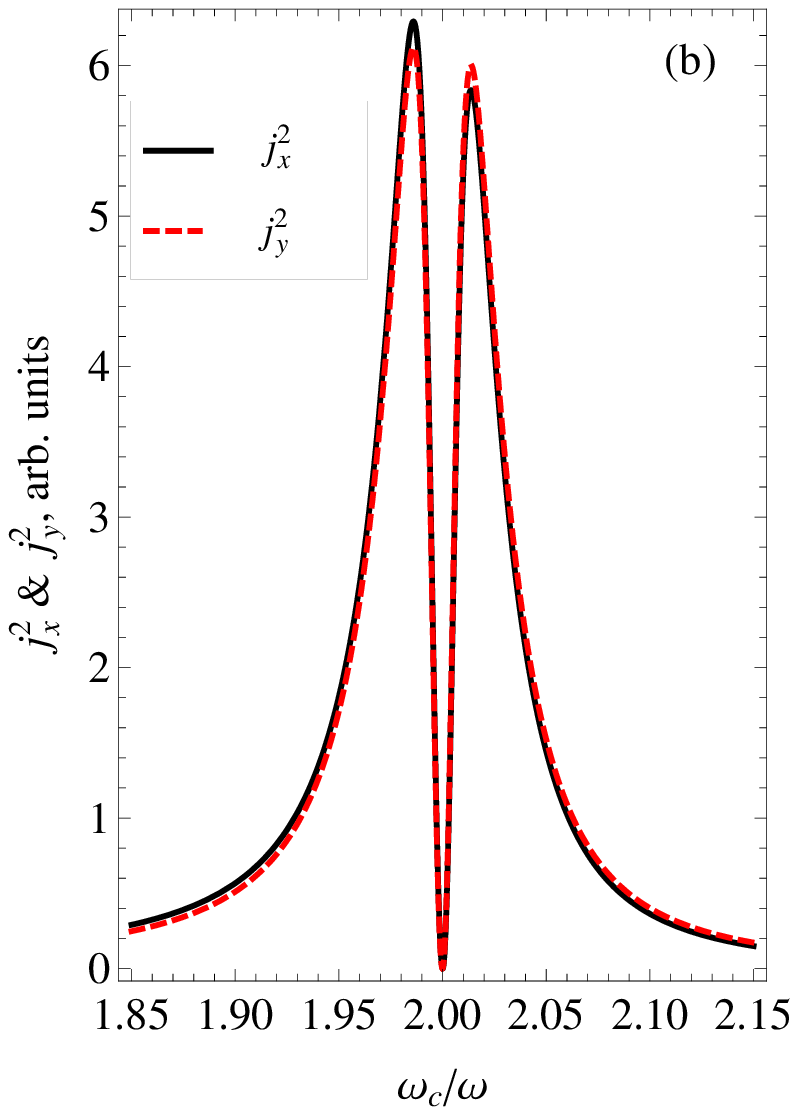}
\caption{\label{fig:intens} The squared amplitudes of the second-harmonic current (proportional to the intensities $J_\perp^{(2)}$ and $J_\parallel^{(2)}$, Eq.  (\ref{intens})) around (a) $\omega_c\simeq \omega$ and (b) $\omega_c\simeq 2\omega$. The parameters are the same as in Figure \ref{fig:zrs} (except $T/E_F$ which is not needed). Black solid curves -- $J_\perp^{(2)}$ (the radiation is polarized perpendicular to the boundary 2D gas -- contact), red dashed curve -- $J_\parallel^{(2)}$ (the radiation is polarized parallel to the boundary 2D gas -- contact). }
\end{figure}

Let us now analyze the $B$-dependencies of the obtained results. The frequency and $B$-dependence of the current amplitudes (\ref{j2av}) has a sharp resonance near $2\omega=\omega_c$, due to Lorentz factors in parenthesis in Eq. (\ref{v2}) and many resonances at the $\omega=k\omega_c$, $k=1,2,\dots$, originating from the $\omega$ and $\omega_c$ dependence of the ponderomotive potential $U_{pm}(x_0)$, Eq. (\ref{pond-pot}). To describe the system response quantitatively we have to specify the shape of the function $E_x(x)$, Eq. (\ref{inhom-field}). Assuming that the 2D gas lies at $x>0$ and that the field $E_x(x)$ varies in space as
\be 
E_x(x)=E_0+(E_c-E_0)e^{-x/l},\ \ x>0,\label{inhom-field-model} 
\ee
where $E_c\gg E_0$, and $l$ is a characteristic length, we get for the function $\epsilon_k(x_0)$, Eq. (\ref{epsdef}):
\be 
\epsilon_k(x_0)=
E_0\delta_{k0}+(E_c-E_0)e^{-x_0/l}(-1)^kI_k(r_c/l) .
\label{epsilonK}
\ee 
Substituting (\ref{epsilonK}) into Eqs. (\ref{pond-pot}), (\ref{densfact}), (\ref{j2av}), and (\ref{intens}), we get the results shown in Figures \ref{fig:zrs}--\ref{fig:intens}. Figure \ref{fig:zrs} shows the ponderomotive potential (\ref{pond-pot}) and the density factor (\ref{densfact}), which determines the observed potential difference $U_{xx}$ between the contacts, see Ref. \cite{Mikhailov11a} and Section \ref{subsec:mirozrs} above. 

Figure \ref{fig:current} exhibits the amplitude of the second-harmonic current (\ref{j2av}). The current has the same oscillations as the ponderomotive potential,  Figures \ref{fig:zrs}(a), and in addition, a resonance near $\omega_c/\omega=2$. Figure \ref{fig:intens} shows the intensity of radiation (\ref{intens})). The intensity of radiation polarized perpendicular to the contact boundary (solid black curves) is close to (at $\omega_c\simeq 2\omega$) or larger than (at $\omega_c\lesssim \omega$) that of the wave polarized parallel to it (red dashed curves). 


Both the MIRO/ZRS and SHG effects have the same origin -- the ponderomotive force effect, but it is important to emphasize the essential difference between their temperature and power dependencies. In order to observe the MIRO, and especially the MIZRS phenomenon, the amplitude of the ponderomotive potential should be comparable with the Fermi energy of electrons, see Eq. (\ref{densfact}). Therefore the amplitude of MIRO oscillations first (when $U_{pm}\lesssim E_F$) grows linearly with the microwave power $P$, and then, in the MIZRS regime $U_{pm}\gtrsim E_F$, depends exponentially on $P$ and $T$, $\ln U_{xx}\propto -P/T$. As a result, the very large resistance oscillations and especially zero resistance states can be observed only at a sufficiently strong input power ($\gtrsim 1$ mW/cm$^2$) and at very low temperatures ($\lesssim 1$ K). In contrast, the second harmonic current amplitudes $j^{(2)}$ are proportional to $U_{pm}$ and almost do not depend on the temperature (a weak $T$-dependence enters the formula (\ref{j2av}) only through the temperature dependence of the scattering rate $\gamma$, i.e. of the mobility). Therefore, the second harmonic intensity should be proportional to the square of the incident wave power and be almost independent of the temperature. Due to the same reasons also the very-high-mobility samples are not vitally important for observation of the SHG effect. These predictions are very important for the experimental observation of the SHG effect, as well as for its practical applications: it should be possible to observe SHG not only at liquid helium  but also at liquid nitrogen or even room temperatures and in samples with a moderate electron mobility.

Let us estimate the expected power of the second-harmonic radiation in the ultra-clean samples in which the MIZRS effect was observed. As seen from Eq. (\ref{j2av}) and Figures \ref{fig:current}--\ref{fig:intens}, the maximum of the second-harmonic current is achieved at $\omega_c-2\omega \simeq \pm \gamma$, so that $|j^{(2)}_{max}|\simeq e n_s|U_{pm}^L|/4m W \gamma$. To be more realistic we will replace here the scattering rate $\gamma$ by a much larger value of the radiative decay $\Gamma$, Eq. (\ref{raddecay}); in this way we assume that the resonances in Figures \ref{fig:current}--\ref{fig:intens} are broader and their heights  are smaller. Then, the absolute value of the ponderomotive potential $U_{pm}^L$ depends on details of the contact and is, strictly speaking, unknown. However, we can use the fact that in the MIZRS regime the value $U_{pm}^L$ exceeds, at least, the Fermi energy, $|U_{pm}^L|\gtrsim E_F$. So we get the following estimates for the second-harmonic current and the corresponding amplitude of the electric field of the emitted second-harmonic wave
\be 
|j^{(2)}_{max}|\gtrsim  
\frac{ e n_sE_F}{4m W \Gamma},\ \ \textrm{and }\ \ 
|E^{(2)}|\gtrsim \frac {\pi e n_sE_F}{2m cW \Gamma}=\frac {E_F}{4 eW}.
\ee
For a typical electron density $ n_s\simeq 3\times 10^{11}$ cm$^{-2}$ and the sample width $W\simeq 50$ $\mu$m we then obtain $E^{(2)}\simeq 0.5$ V/cm and $J^{(2)}\simeq 0.8$ mW/cm$^2$. The estimated power density of the emitted second harmonic turns out to be comparable with the power of the incident wave. Even if this estimate seems to be too optimistic (although we really kept all parameters within  realistic boundaries), it definitely shows that the predicted effect is not small, should be easily observed in  high-quality two-dimensional electron systems, and has a great potential for electronic applications.

\section{Conclusions\label{sec:conclusion}}

The great advances of technology \cite{Pfeiffer03}, which led to the improvement of quality of GaAs quantum well samples by many orders of magnitude in the past decades, resulted in the discoveries of many interesting and intriguing physical effects including the MIRO/ZRS phenomena. A practically collisionless solid-state electron plasma has been in fact created. A complete understanding of the MIRO/ZRS phenomena, achieved with the development of the ponderomotive-forces theory, and the prediction of the second-harmonic generation in such systems open up new perspectives in studying \textit{nonlinear electrodynamic} phenomena, such as plasma instabilities, frequency multiplication, frequency mixing effects, etc., in such collisionless two-dimensional electron plasma. This has a great potential for practical applications of the discussed effects in microwave and terahertz electronics.
 
\begin{acknowledgments}
I would like to thank Michel Dyakonov, Ramesh Mani, Stefan Wiedmann and Vladimir Volkov for useful discussions of the MIRO/ZRS effects. Special thanks must be given to Alexei Chepelianskii who performed numerical calculations of the influence of the depletion layers on the two-terminal resistance of the Hall bar (Section \ref{subsubsec:two-terminalresistance}). I also thank the Deutsche Forschungsgemeinschaft for financial support of this work. 
\end{acknowledgments}

\bibliography{../../../BIB-FILES/lowD,../../../BIB-FILES/fqhe,../../../BIB-FILES/mikhailov,../../../BIB-FILES/zerores}
\end{document}